\documentclass[%
reprint,
superscriptaddress,
 amsmath,amssymb,
 aps,
prb,
longbibliography
]{revtex4-2}

\usepackage{tikz}
\usepackage{graphicx}% Include figure files
\usepackage{dcolumn}% Align table columns on decimal point
\usepackage{bm}% bold math
\usepackage[section]{placeins}
\usepackage{braket}
\usepackage{xcolor}
\usepackage{hyperref}
\hypersetup{
    colorlinks=true,
    linkcolor=blue,   % Equations, figures, sections
    citecolor=blue,  % Citations
    urlcolor=blue    % URLs
}
\usepackage{enumitem}

\makeatletter
\def\@fnsymbol#1{\ensuremath{\ifcase#1\or \dagger\or \ddagger\or \mathsection\or \mathparagraph\or \|\or **\or \dagger\dagger\or \ddagger\ddagger\else\@ctrerr\fi}}
\makeatother

\newcommand{\PBA}{Poly-bi-[8]-annulenylene}
\newcommand{\pba}{poly-bi-[8]-annulenylene}

\begin{document}

\preprint{APS/123-QED}

\title{Symmetry-Controlled Quantum State Transfer in Poly-bi-[8]-annulenylene: From Heisenberg to Hubbard Dynamics}

\author{Adnan Jaljuli}
\affiliation{Department of Computer Science, NC State University, Raleigh, NC, USA}

\author{Pablo Serra}
\affiliation{Instituto de Física Enrique Gaviola (CONICET-UNC) and Facultad de Matemática, Astronomía, Física y Computación, Universidad Nacional de Córdoba, Córdoba, Argentina}
\affiliation{Department of Electrical and Computer Engineering, NC State University, Raleigh, NC, USA}

\author{Ammar Ali}
\affiliation{Department of Electrical and Computer Engineering, NC State University, Raleigh, NC, USA}

\author{Sabre Kais}
\email{skais@ncsu.edu}
\affiliation{Department of Electrical and Computer Engineering, NC State University, Raleigh, NC, USA}
\affiliation{Department of Chemistry and Department of Physics, NC State University, Raleigh, NC, USA}

\date{\today}

\begin{abstract}
We investigate quantum state transfer in the poly-bi-[8]-annulenylene lattice using isotropic Heisenberg and Hubbard models. For a 16-site lattice, reflection symmetry separates the single-excitation Hamiltonian into independent triplet and singlet sectors, allowing analytical determination of the energy spectra. Although perfect state transfer is forbidden for positive couplings, appropriately chosen coupling ratios enable pretty-good transfer with arbitrarily high fidelity. Remarkably, the triplet transfer probability is independent of the interchain rung coupling, while singlet transfer can be selectively enhanced or suppressed, enabling symmetry-selective quantum transport without time-dependent control. Extending the analysis to the Hubbard model, we show that the Heisenberg transfer dynamics emerge in the strong-interaction limit, while finite interactions delay all high-fidelity transfer peaks of a symmetry sector by the same fraction of their transfer time, set by the spectrum, and reduce the peak fidelity, even at the corrected time, by an amount that grows with the transfer time. These findings are relevant to future quantum-network architectures, where symmetry-controlled state transfer could provide a building block for connecting quantum memories and emitter-based interfaces within quantum nodes. Overall, our results demonstrate how molecular network geometry and symmetry can be exploited to engineer and control quantum-state transport in interacting spin systems, providing a theoretical framework for quantum information transfer in molecular-inspired quantum architectures.

\end{abstract}

\maketitle

\section{\label{sec:level1}Introduction}
Transferring quantum states between distant parts of a quantum device is a
basic requirement for scalable quantum information processing. Bose proposed
doing this with a chain of permanently coupled spins~\cite{boseQuantumCommunicationUnmodulated2003}: the
state propagates under the natural, time-independent dynamics of the chain,
and no control is needed during the transfer. This idea has grown into a broad
field that studies how the geometry and couplings of a spin network determine
its quality as a quantum channel~\cite{kayPERFECTEFFICIENTSTATE2010}, and state transfer has been
demonstrated experimentally, for example in quantum-dot spin chains coupled by
Heisenberg exchange~\cite{kandelCoherentSpinstateTransfer2019a}.

The quality of a channel is measured by the transfer probability
$P(\tau)=|\langle b|e^{-iH\tau}|a\rangle|^{2}$ from a source state $|a\rangle$ to a
target state $|b\rangle$ at time $\tau$. Perfect transfer (PT), $P(\tau)=1$ at a finite time,
can be achieved in chains with engineered
couplings~\cite{christandlPerfectStateTransfer2004a}.
In pretty good transfer (PGT), $P(\tau)$ comes arbitrarily close to one at
suitable times~\cite{godsilNumberTheoreticNatureCommunication2012,banchiPrettyGoodState2017a}. 
Most of this theory concerns XX interactions, for which the single-excitation
Hamiltonian is the adjacency matrix of the coupling graph. In the isotropic
Heisenberg (XXX) exchange, natural for electron spins in quantum dots and
molecules~\cite{kandelCoherentSpinstateTransfer2019a,gaita-arinoMolecularSpinsQuantum2019a}, 
each site
acquires an on-site energy set by the total strength of its bonds. In uniform Heisenberg chains PGT between the ends
requires the number of sites to be a power of two~\cite{banchiPrettyGoodState2017a}, unlike
the XX case. PGT also
says nothing about when high fidelities are reached, so PGT alone is of limited practical use.
It is more relevant to ask whether the error can be kept below a tolerance at a
prescribed, experimentally accessible time by tuning a few
couplings~\cite{serraPrettyGoodQuantum2022}; we call this
almost perfect transfer (APT).

Networks with several paths, such as parallel chains~\cite{burgarthConclusiveArbitrarilyPerfect2005} and
spin ladders~\cite{pushpanQuantumStateTransfer2024}, offer further possibilities. When the
network has a reflection symmetry, the dynamics splits into symmetric and
antisymmetric sectors, a reduction used to analyze PGT between mirror-image
sites~\cite{kemptonPrettyGoodQuantum2017}. Chains with alternating couplings, as in the
Su--Schrieffer--Heeger (SSH) model~\cite{suSolitonExcitationsPolyacetylene1980a}, host
boundary zero modes in their topological phase, which have been used for
robust transfer
protocols~\cite{estarellasTopologicallyProtectedLocalised2017,meiRobustQuantumState2018a},
usually at the cost of long transfer times or time-dependent control. These edge
states are not required for high-fidelity transfer, however, and SSH-type XXZ
chains show none near the isotropic Heisenberg point~\cite{moraguesAnisotropicHeisenbergSuSchriefferHeeger2026a}.

\PBA{}, a polymer of cyclooctatetraene units, combines these features: its
two arms have SSH-like alternating couplings and are joined by rungs, giving
it a reflection symmetry and two propagation paths. In a tight-binding
description it behaves, for weak coupling between the arms, as two coupled
SSH chains with boundary zero
modes~\cite{muruganandamForayTopologyPolybi82023a}. 
Motivated by the growing
role of molecular systems in quantum information
science~\cite{gaita-arinoMolecularSpinsQuantum2019a}, we ask whether this connectivity can serve as
an efficient channel when the sites interact through isotropic Heisenberg
exchange. Because the spin model adds the on-site terms described above,
tight-binding results, including those on edge states, do not carry over automatically.

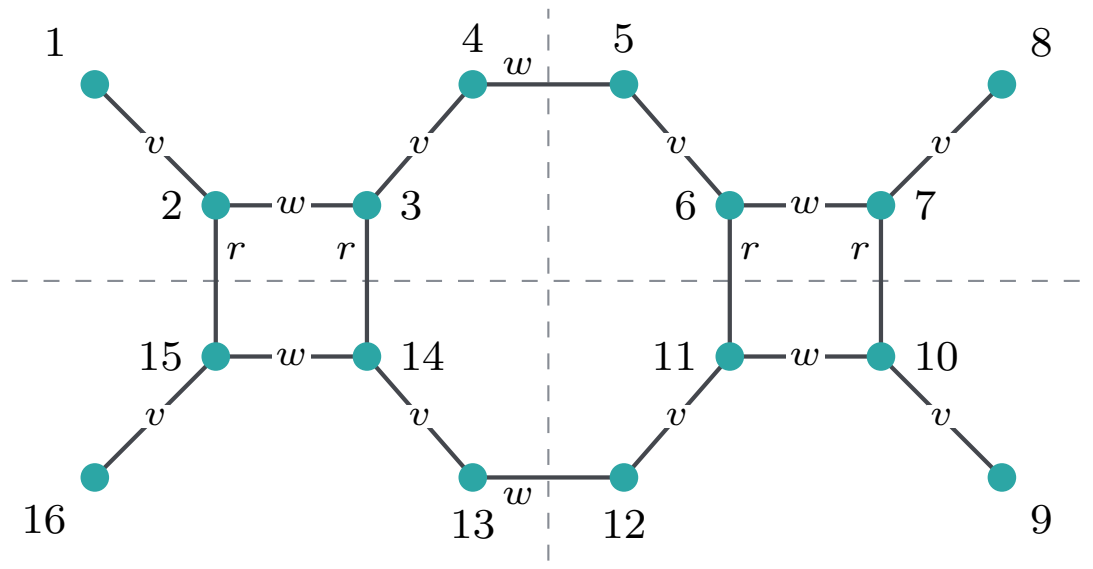
\begin{figure}[!tbp]
\centering
\definecolor{figSite}{HTML}{2CA6A5}
\definecolor{figBond}{HTML}{44474D}
\definecolor{figAxis}{HTML}{878C94}
\begin{tikzpicture}[scale=1.0,
  A/.style={circle,fill=figSite,inner sep=1.9pt},
  B/.style={circle,fill=figSite,inner sep=1.9pt},
  bond/.style={thick,draw=figBond},
  axis/.style={dashed,draw=figAxis},
  lab/.style={font=\footnotesize},
  cpl/.style={font=\scriptsize\itshape,fill=white,inner sep=0.6pt},
  cpls/.style={font=\scriptsize\itshape,inner sep=1.5pt}]
  \coordinate (n1) at (-0.8,1.8);  \coordinate (n2) at (0,1);    \coordinate (n3) at (1,1);
  \coordinate (n4) at (1.7,1.8);   \coordinate (n5) at (2.7,1.8); \coordinate (n6) at (3.4,1);
  \coordinate (n7) at (4.4,1);     \coordinate (n8) at (5.2,1.8);
  \coordinate (n9) at (5.2,-0.8);  \coordinate (n10) at (4.4,0);  \coordinate (n11) at (3.4,0);
  \coordinate (n12) at (2.7,-0.8); \coordinate (n13) at (1.7,-0.8);\coordinate (n14) at (1,0);
  \coordinate (n15) at (0,0);      \coordinate (n16) at (-0.8,-0.8);
  \draw[axis] (-1.35,0.5) -- (5.75,0.5);  % horizontal: exchanges the arms, j <-> 17-j (triplet/singlet)
  \draw[axis] (2.2,-1.35) -- (2.2,2.3);   % vertical: exchanges the ends
  % v bonds (diagonal)
  \foreach \a/\b in {n1/n2,n3/n4,n5/n6,n7/n8,n9/n10,n11/n12,n13/n14,n15/n16}
     \draw[bond] (\a) -- node[cpl]{$v$} (\b);
  % w bonds inside the squares
  \foreach \a/\b in {n2/n3,n6/n7,n10/n11,n14/n15}
     \draw[bond] (\a) -- node[cpl]{$w$} (\b);
  % w bonds on the octagon (labels off the vertical axis)
  \draw[bond] (n4) -- node[cpls,above,pos=0.3]{$w$} (n5);
  \draw[bond] (n13) -- node[cpls,below,pos=0.3]{$w$} (n12);
  % r bonds (labels inside the squares, above the horizontal axis)
  \draw[bond] (n2) -- node[cpls,right,pos=0.3]{$r$} (n15);
  \draw[bond] (n3) -- node[cpls,left,pos=0.3]{$r$}  (n14);
  \draw[bond] (n6) -- node[cpls,right,pos=0.3]{$r$} (n11);
  \draw[bond] (n7) -- node[cpls,left,pos=0.3]{$r$}  (n10);
  % sites (both sublattices drawn alike)
  \node[A,label={[lab]above left:1}]   at (n1)  {};
  \node[B,label={[lab]left:2}]         at (n2)  {};
  \node[A,label={[lab]right:3}]        at (n3)  {};
  \node[B,label={[lab]above:4}]        at (n4)  {};
  \node[A,label={[lab]above:5}]        at (n5)  {};
  \node[B,label={[lab]left:6}]         at (n6)  {};
  \node[A,label={[lab]right:7}]        at (n7)  {};
  \node[B,label={[lab]above right:8}]  at (n8)  {};
  \node[A,label={[lab]below right:9}]  at (n9)  {};
  \node[B,label={[lab]right:10}]       at (n10) {};
  \node[A,label={[lab]left:11}]        at (n11) {};
  \node[B,label={[lab]below:12}]       at (n12) {};
  \node[A,label={[lab]below:13}]       at (n13) {};
  \node[B,label={[lab]right:14}]       at (n14) {};
  \node[A,label={[lab]left:15}]        at (n15) {};
  \node[B,label={[lab]below left:16}]  at (n16) {};
\end{tikzpicture}
    \caption{Poly-bi-[8]-annulenylene with $N=2$,  with coupling parameters $(v,w,r)$. The horizontal dashed line is the reflection axis that exchanges the two arms, $j\leftrightarrow17-j$, and defines the triplet and singlet sectors (Sec.~\ref{sec:basis}). The vertical dashed line is the reflection axis that exchanges the ends; each subgraph to either side of it can be considered its own \pba{} model with $N=1$.  }
    \label{fig:pb8a_N2}
\end{figure}

We treat in detail the structure with $N=2$ unit cells hosting 16 sites
(Fig.~\ref{fig:pb8a_N2}). Our main results are:
\begin{enumerate}[label=(\roman*), leftmargin=*, itemsep=1pt, topsep=3pt]
\item Reflection symmetry splits the single-excitation Hamiltonian into two
  $8\times 8$ blocks, so transfer reduces to moving a triplet or singlet state
  from the end pair $(1,16)$ to the opposite pair $(8,9)$. All eigenenergies
  follow in closed form. For any two-leg ladder, the rung coupling $r$ only
  shifts the triplet sector by a constant, so triplet transfer is that of two
  uncoupled staggered chains~\cite{serraExactSolutionFamily2022}.
\item No positive couplings give PT in either sector: in the triplet
  sector the transition frequencies are incommensurate, and in the singlet
  sector the PT conditions are incompatible modulo~4.
\item Couplings related to Pythagorean triples leave only two rationally
  independent irrational frequencies and give PGT of both states. The arrival
  times, bounded with the Lenstra--Lenstra--Lov\'asz
  algorithm~\cite{lenstraFactoringPolynomialsRational1982}, follow the scaling
  law of Ref.~\cite{serraScalingLawArrival2023a}. PGT also holds between
  mirror-image bulk sites~\cite{banchiPrettyGoodState2017a}.
\item Because the triplet sector is independent of $r$, the rung coupling can
  tune the singlet transfer probability at a fixed time close to one or zero
  without changing the triplet transfer. Both states can then be transferred
  with high fidelity, or an unwanted singlet component can be suppressed.
\item The rungs leave the clean triplet transfer unchanged but make it more
  robust to static coupling disorder than a single SSH-type chain.
\item An out-of-time-order correlator
  (OTOC)~\cite{xuScramblingDynamicsOutofTimeOrdered2024a} built from
  projectors, as in fidelity
  OTOCs~\cite{lewis-swanUnifyingScramblingThermalization2019a}, equals
  $P(\tau)^{2}$ and carries no information beyond the transfer probability.
\end{enumerate}

To test how charge fluctuations modify these predictions, we extend the
analysis to the half-filled Hubbard model. The Heisenberg transfer
probabilities are recovered at strong on-site repulsion $U$. At large but
finite $U$, the high-fidelity peaks acquire a relative delay proportional to
$(t_v/U)^2$, where $t_v$ is the hopping amplitude on the $v$ bonds. We show
that, at fixed couplings and initial state, this prefactor is the same for all
high-fidelity peaks within a sector: it is a spectral average of the finite-$U$
shifts of the transition frequencies and does not depend on the peak. The loss
that remains after correcting the readout time grows with the transfer time, so
the early peaks retain high transfer probabilities at moderate $U/t_v$, whereas the late PGT
peaks require much stronger repulsion.

The paper is organized as follows: Sec.~\ref{sec:model} introduces the model,
Sec.~\ref{sec:transfer} studies PT, PGT, and APT and their robustness to
coupling disorder, Sec.~\ref{sec:hubbard} examines transfer in the Hubbard
model and its approach to the Heisenberg limit, and
Sec.~\ref{sec:conclusion} concludes.
Appendix~\ref{app:rindep} shows that the triplet transfer is independent
of the rungs for any two-leg ladder,
Appendix~\ref{app:ptexplicit} gives the explicit transfer probability,
Appendix~\ref{app:otoc} discusses OTOCs, and Appendix~\ref{app:delay}
derives the delay of the transfer peaks at finite $U$.

\section{\label{sec:model} Heisenberg hamiltonian, initial states, eigenenergies}

\subsection{The \PBA{} lattice}
A structure of $N$ unit cells has $8N$ sites and three exchange couplings:
$v$ and $w$ alternate along each of the two arms, as in the SSH
chain~\cite{suSolitonExcitationsPolyacetylene1980a}, and $r$ couples the two
arms.
For weak $r$ and $v<w$, the corresponding tight-binding model hosts boundary
zero modes under open boundary
conditions~\cite{muruganandamForayTopologyPolybi82023a}.

We number the sites $j=1,\dots,8N$ by following one arm from end to end and
then returning along the other, so that $j=1,\dots,4N$ lie on one arm and
$j=4N+1,\dots,8N$ on the other. The bonds are then
\begin{equation}
\begin{aligned}
v &: \; (j,\,j{+}1), && j \ \text{odd}, \\
w &: \; (j,\,j{+}1), && j \ \text{even}, \\
r &: \; (j,\,8N{+}1{-}j), && j \le 4N , \; j \equiv 2,3 \;(\mathrm{mod}\ 4),
\end{aligned}
\label{eq:bonds}
\end{equation}
with $j$ and $j{+}1$ on the same arm in the first two lines, giving $4N$ bonds
of strength $v$, $4N-2$ of strength $w$ and $2N$ rungs. We treat $N=2$ (16 sites, Fig.~\ref{fig:pb8a_N2}) in detail; the symmetry
decomposition and the rung independence of the triplet sector hold for any $N$.

\subsection{\label{subsec:heisenberg_ham}Heisenberg Hamiltonian}

We look at the spin-$\frac{1}{2}$ isotropic Heisenberg (XXX) ferromagnetic model in the single-excitation sector.

For a general $N$, the Hamiltonian is
\begin{equation}
\begin{split}
H_N = &-v\sum_{\substack{j=1\\ j\ \text{odd}}}^{8N-1} \bm{\sigma}_j\cdot\bm{\sigma}_{j+1}\\
&- w\sum_{\substack{j=2\\ j\ \text{even},\ j\neq 4N}}^{8N-2} \bm{\sigma}_j\cdot\bm{\sigma}_{j+1} \\
&- r\sum_{\substack{j=1\\ j\equiv 2,3\ (\mathrm{mod}\ 4)}}^{4N} \bm{\sigma}_j\cdot\bm{\sigma}_{8N+1-j}
\end{split}
\label{eq:HN}
\end{equation}

\noindent where $\bm{\sigma}_i = (\sigma_i^x,\sigma_i^y, \sigma_i^z)$. We set $\hbar = 1$ and measure the couplings $v$, $w$, $r$ in units of an
arbitrary energy scale $J_0$, so that the time $\tau$ is measured in units of
$\hbar/J_0$. We use $\tau$ for time throughout, since $t$ denotes the hopping amplitudes of the
Hubbard model in Sec.~\ref{sec:hubbard}. Rescaling all couplings by a factor $\lambda$ rescales the
dynamics as $\tau \to \tau/\lambda$, so the transfer probabilities depend only on
the ratios of the couplings and on the products of couplings and time. 

The Hamiltonian in Eq.~\eqref{eq:HN} commutes with the total magnetization along $z$,

\begin{equation}
\label{erchm}
 \left[ H_N, \sum_i \sigma_i^z \right] = 0,
\end{equation}
\noindent therefore, the Hamiltonian block-diagonalizes in subspaces of fixed number of spins up. For our initial state, we consider a normalized single-excitation state for the spin model. The $8N$ states with a single spin up are denoted as follows
\begin{equation}
|1 \rangle = |100\ldots 0\rangle, |2 \rangle =
|010\ldots0\rangle,\ldots , |8N \rangle = |000\ldots 1\rangle.
\end{equation}

% ###################################################

\subsection{\label{sec:basis} Singlet and triplet basis, and the single-excitation sector}
The poly-bi-[8]-annulenylene molecule is symmetric under reflection about both the horizontal and the vertical axes. Since we are interested in information transfer from left to right, we use the horizontal reflection symmetry to block-diagonalize the Hamiltonian matrix into two $4N \times 4N$ blocks. This symmetry also has a clear physical interpretation: in this new basis, we study not the transfer of an individual spin up from site 1 to site 8 or 9, but rather the transfer of a triplet or singlet state from the pair of sites (1,16) to the pair of sites (8,9).
The basis states with triplet and singlet symmetry are
\begin{equation}
\label{eq:pair-states}
|j_{\pm}\rangle = \frac{1}{\sqrt{2}}\left ( |j\rangle \pm |8N+1-j\rangle \right),
\qquad j = 1, \dots, 4N
\end{equation}

In the single-excitation subspace, and for $N=2$, the Hamiltonian can be written as two blocks of  $8 \times 8$ matrices, one for the triplet sector and one for the singlet, i.e.

\begin{equation}
H_2 =
\left(\begin{array}{c|c}
  \underbrace{h^{(+)}}_{8\times 8} & \mathbf{0} \\ \hline
  \mathbf{0} & \underbrace{h^{(-)}}_{8\times 8}
\end{array}\right)
= h^{(+)} \oplus h^{(-)}.
\end{equation}

With the shorthand $a \equiv 3v+3w+2r$, $b \equiv 3v+2w+2r$ and $c \equiv 3v+2w$,
the triplet block reads
\begin{equation}
\label{emh8t}
h^{(+)} = -2
\begin{pmatrix}
 a & v \\
 v & b & w \\
   & w & b & v \\
   &   & v & b & w \\
   &   &   & w & b & v \\
   &   &   &   & v & b & w \\
   &   &   &   &   & w & b & v \\
   &   &   &   &   &   & v & a
\end{pmatrix},
\end{equation}
and the singlet block is
\begin{equation}
\label{emh8s}
h^{(-)} = -2
\begin{pmatrix}
 a & v \\
 v & c & w \\
   & w & c & v \\
   &   & v & b & w \\
   &   &   & w & b & v \\
   &   &   &   & v & c & w \\
   &   &   &   &   & w & c & v \\
   &   &   &   &   &   & v & a
\end{pmatrix}.
\end{equation}

\subsection{\label{sec:eigen} Eigenenergies and spectral weights of eigenstates}
The eigenenergies corresponding to the triplet-symmetry eigenvectors are
\begin{align}
\label{eenertri}
&E_1^{(+)}=-2 (2 v +3 w+2 r)\nonumber \\
&E_2^{(+)}=-2 (4 v +3 w+2 r)\nonumber \\
&E_3^{(+)}=-2 (3 v +2 w+2 r-\sqrt{v^2+w^2})\nonumber \\
&E_4^{(+)}=-2 (3 v +2 w+2 r+\sqrt{v^2+w^2}) \nonumber \\
&E_5^{(+)}=-2 (3 v +2 w+2 r-\sqrt{v^2+w^2-\sqrt{2} v w}) \nonumber \\
&E_6^{(+)}=-2 (3 v +2 w+2 r+\sqrt{v^2+w^2-\sqrt{2} v w})
\nonumber \\
&E_7^{(+)}=-2 (3 v +2 w+2 r-\sqrt{v^2+w^2+\sqrt{2} v w}) \nonumber \\
&E_8^{(+)}=-2 (3 v +2 w+2 r+\sqrt{v^2+w^2+\sqrt{2} v w})
\end{align}

And the eigenenergies associated with the singlet-symmetry eigenvectors are
\begin{align}
\label{eenersing}
&E_1^{(-)}=-2 (3 v +3 w+r-\sqrt{v^2+r^2}) \nonumber \\
&E_{2}^{(-)}=-2 (3 v +3 w+r+\sqrt{v^2+r^2}) \nonumber \\
&E_{3}^{(-)}=-2 (3 v +2 w+r-\sqrt{v^2+(w+r)^2}) \nonumber \\
&E_{4}^{(-)}=-2 (3 v +2 w+r+\sqrt{v^2+(w+r)^2}) \nonumber \\
&E_{5}^{(-)}=-2 \Big(3 v +2 w+r \nonumber \\
&\hphantom{E_{5}^{(-)}=-2 \Big(}{}-\sqrt{v^2+w^2+r^2-w\sqrt{2 (v^2 +2 r^2)}}\,\Big) \nonumber \\
&E_{6}^{(-)}=-2 \Big(3 v +2 w+r \nonumber \\
&\hphantom{E_{6}^{(-)}=-2 \Big(}{}+\sqrt{v^2+w^2+r^2-w\sqrt{2 (v^2 +2 r^2)}}\,\Big) \nonumber \\
&E_{7}^{(-)}=-2 \Big(3 v +2 w+r \nonumber \\
&\hphantom{E_{7}^{(-)}=-2 \Big(}{}-\sqrt{v^2+w^2+r^2+w\sqrt{2 (v^2 +2 r^2)}}\,\Big) \nonumber \\
&E_{8}^{(-)}=-2 \Big(3 v +2 w+r \nonumber \\
&\hphantom{E_{8}^{(-)}=-2 \Big(}{}+\sqrt{v^2+w^2+r^2+w\sqrt{2 (v^2 +2 r^2)}}\,\Big)
\end{align}

In the triplet energies of Eq.~\eqref{eenertri}, $r$ appears only through
the common shift $-4r$ of every eigenvalue, whereas in the singlet
energies of Eq.~\eqref{eenersing} it enters the square roots.
Appendix~\ref{app:rindep} shows that the first property holds for any
number of unit cells and even for site-dependent rung couplings: the rungs
add a multiple of the identity to the triplet block, so its eigenvectors,
and with them the triplet transfer probability, do not depend on $r$.

\begin{figure}
    \centering
    \includegraphics[width=1\linewidth]{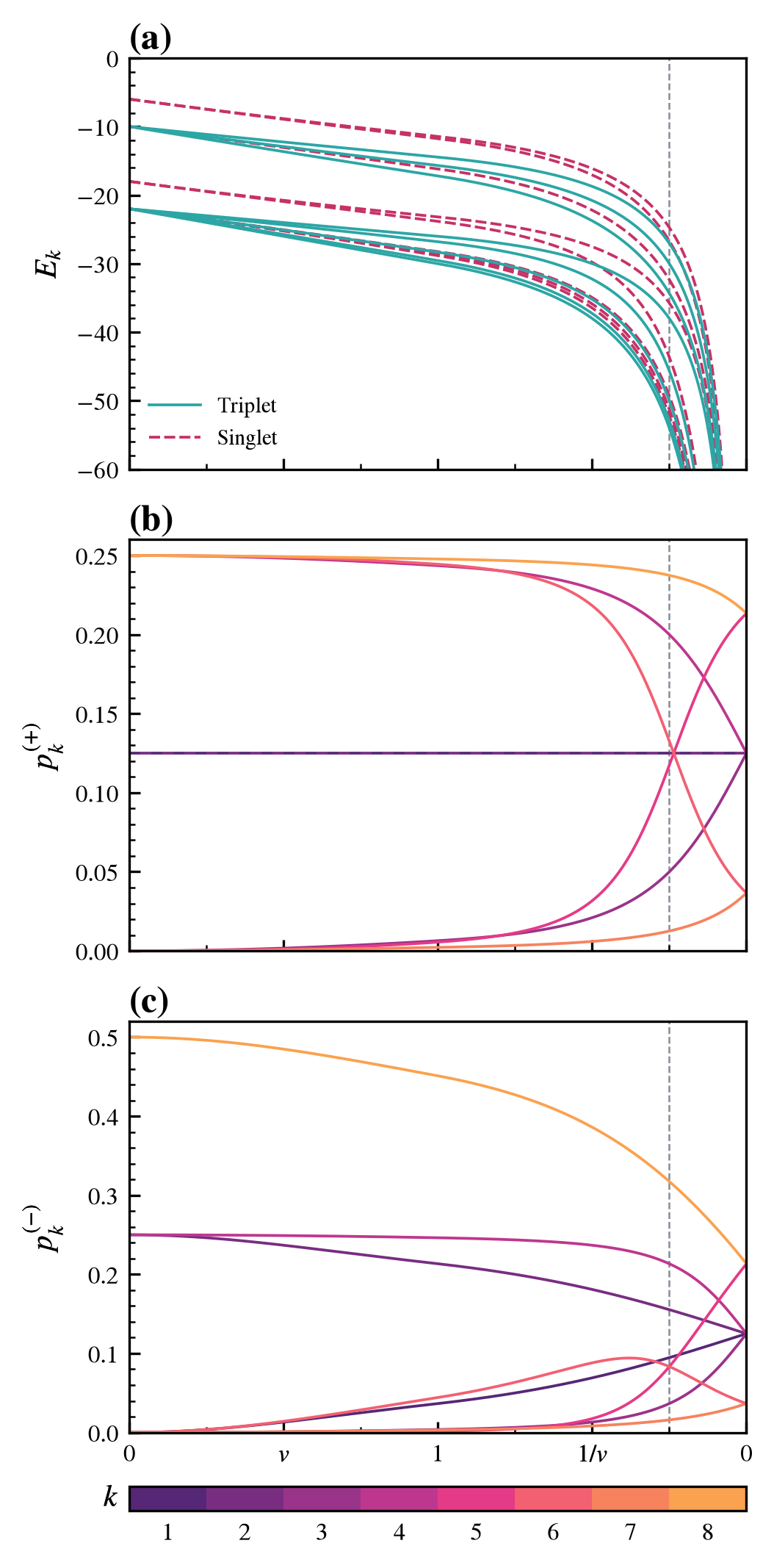}
    \caption{(a) Energy spectrum for $w=3$ and $r=1$. Teal (rose) lines correspond to eigenstates with triplet (singlet) symmetry. \\
    Panels (b) and (c) show the spectral weights $p_k=|\langle 1_\pm|\phi_k\rangle|^2$ of the 
    % eight
triplet (b) and singlet (c) eigenvectors, for $w=3$ and $r=1$; $k$ labels the
eigenvalues as in Eqs.~\eqref{eenertri} and~\eqref{eenersing}. By mirror symmetry, the weights on $|8_\pm\rangle$ are the
same. \\ The $x$ axes run over $v\in[0,1]$, then over $1/v$ from 1 to 0. The
vertical dashed line marks $v=4$, the value used for the triplet example of Sec.~\ref{sec:transfer}.}
\label{fig:eigen_energies_and_loc}
\end{figure}

The energies for the particular case $w=3$ and $r=1$ are shown in Fig.~\ref{fig:eigen_energies_and_loc}.
We denote by $|\phi_k\rangle$ the eigenvector with eigenvalue $E_k$ of
Eq.~\eqref{eenertri} or~\eqref{eenersing}, and by
$p_k=|\langle 1_\pm|\phi_k\rangle|^2$ its spectral weight on the initial
state. The eigenvectors, and with them the weights, follow analytically from
Eqs.~\eqref{emh8t} and~\eqref{emh8s} but are lengthy, and we do not list
them. Figure~\ref{fig:eigen_energies_and_loc} shows the weights; by mirror symmetry, those
on $|8_\pm\rangle$ are the same.

\section{\label{sec:transfer}Perfect, pretty good, and almost perfect transfer in the Heisenberg model}

The Hamiltonian is block diagonal in the basis of Eq.~\eqref{eq:pair-states},
so a state prepared in one sector stays there. We therefore treat two
independent problems: the transfer of the triplet state $|1_+\rangle$ to
$|8_+\rangle$ under $h^{(+)}$ [Eq.~\eqref{emh8t}], and of the singlet
state $|1_-\rangle$ to $|8_-\rangle$ under $h^{(-)}$ [Eq.~\eqref{emh8s}].
Their transfer probabilities are
\begin{equation}
\label{eP}
P^{(\pm)}(\tau)=\left|\langle 8_\pm|\,e^{-ih\tau}|1_\pm\rangle\right|^2 ,
\end{equation}
and we drop the superscript for statements that hold in both sectors. For
transfer between other pairs of Eq.~\eqref{eq:pair-states} we write
$P^{(\pm)}_{j\to k}(\tau)=|\langle k_\pm|e^{-ih\tau}|j_\pm\rangle|^2$, so that
$P^{(\pm)}(\tau)=P^{(\pm)}_{1\to 8}(\tau)$.

The network exhibits PT if $P(\tau_{\rm PT})=1$ at some finite time
$\tau_{\rm PT}$~\cite{kayPERFECTEFFICIENTSTATE2010,christandlPerfectStateTransfer2004a},
and PGT if the transfer probability gets arbitrarily close to one, even if it
never reaches it~\cite{banchiPrettyGoodState2017a,godsilNumberTheoreticNatureCommunication2012}:
\begin{equation}
\label{edpgt}
\forall \epsilon >0 \;\exists \;\tau_\epsilon :  P(\tau_\epsilon)\geq1-\epsilon\,.
\end{equation}
Following Ref.~\cite{serraScalingLawArrival2023a}, we call the earliest time
at which $P(\tau)\ge 1-\epsilon$ the arrival time $\tau_\epsilon$. APT, the
finite-time criterion introduced in Sec.~\ref{sec:level1}, is made precise in
Sec.~\ref{sec:apt}.

Consider either of the two $8\times8$ blocks, denoted by $h$.
Each is a real symmetric tridiagonal matrix with nonzero
nearest-neighbor off-diagonal elements. Its eigenvalues $E_k$ are
therefore nondegenerate, and every eigenvector $|\phi_k\rangle$
has nonzero amplitudes at both ends~\cite{parlettSymmetricEigenvalueProblem1998}.
Each block is also mirror symmetric,
$h_{jk}=h_{9-j,\,9-k}$. This reflection exchanges the ends within
each block and is distinct from the reflection that defines the
triplet and singlet sectors. Each eigenvector consequently has a
definite parity $s_k=\pm1$ under $j\to9-j$, with successive
eigenvectors in energy order having opposite
parities~\cite{kayPERFECTEFFICIENTSTATE2010,vinetAlmostPerfectState2012}.

The endpoint amplitudes therefore satisfy
$\langle8_\pm|\phi_k\rangle=s_k\langle1_\pm|\phi_k\rangle$,
so the transfer amplitude within the chosen sector is
\begin{equation}
\label{eamp}
\langle 8_\pm|e^{-ih\tau}|1_\pm\rangle
=\sum_{k=1}^{8}s_k\,p_k\,e^{-iE_k\tau},
\end{equation}
where $p_k=|\langle1_\pm|\phi_k\rangle|^2>0$ and
$\sum_k p_k=1$. This is a weighted sum of unit complex numbers
$s_k e^{-iE_k\tau}$. Since the weights are positive and sum to one,
the amplitude has unit magnitude if and only if all these numbers
have the same phase. Thus $P(\tau)=1$ if and only if
\begin{equation}
\label{ephase}
s_k\,e^{-iE_k\tau}=e^{i\theta}\qquad\text{for all }k
\end{equation}
for some global phase $\theta$. Within the chosen sector, the
mirror-even eigenvectors must therefore acquire a common dynamical
phase, and the mirror-odd eigenvectors must acquire the opposite
phase~\cite{kayPERFECTEFFICIENTSTATE2010}.
PT requires this condition to hold exactly at some finite time,
whereas PGT requires it to be approached arbitrarily closely at
suitable times. The transfer probability can also be written as
a sum of cosines whose frequencies are the eigenvalue differences
$E_k-E_l$.

% ---- 3. Why minimize the number of independent irrationals ------------
As we show in Sec.~\ref{sec:pt}, neither sector admits PT, so the relevant
question is how quickly Eq.~\eqref{ephase} can be approached. Rational
frequencies can all be matched at periodic times, but each rationally
independent irrational number adds a phase that must reach its target
together with the others. As a result, the arrival time grows
as
\begin{equation}
\label{este}
\tau_\epsilon \sim \epsilon^{-K_I/2},
\end{equation}
where $K_I$ is the number of rationally independent irrational numbers
among the frequencies $E_k-E_l$~\cite{serraScalingLawArrival2023a}. Couplings that make PGT practical are
therefore those that minimize $K_I$.

% ---- 4. Pythagorean couplings ------------------------------------------
The eigenvalues of Eqs.~\eqref{eenertri} and \eqref{eenersing} contain
square roots, and $K_I$ is reduced by making as many of them rational as
possible, i.e.\ by choosing couplings that are legs of Pythagorean
triples. The conditions differ between the sectors because the rungs act
differently. 

In the triplet sector $r$ only shifts the spectrum (Sec.~\ref{sec:eigen}),
so $P^{(+)}(\tau)$ is that of two uncoupled eight-site
chains~\cite{serraExactSolutionFamily2022}, and the only removable root is
$\sqrt{v^2+w^2}$, which requires $(v,w)$ to be legs of a Pythagorean triple;
the nested roots $X_\pm=\sqrt{v^2+w^2\pm\sqrt{2}\,vw}$ are irrational for any
rational couplings.
In the singlet sector the rungs act as an on-site potential
$4r$ on sites 2, 3, 6 and 7, so $r$ enters the roots $\sqrt{v^2+r^2}$ and
$\sqrt{v^2+(w+r)^2}$, and the conditions are on $(v,r)$ and $(v,w+r)$.
Satisfying all conditions at once requires large couplings, the smallest
being $(v,w,r)=(60,11,80)$. Since $P^{(+)}$ does not depend on $r$, we
instead choose the couplings separately for each sector.

% ---- 5. Eigenvalues of the two examples ---------------------------------
For the triplet we take $(v,w)=(4,3)$ and $r=1$; other values of $r$
shift every eigenvalue by $-4(r-1)$. In increasing order the eigenvalues
are
\begin{align}
\label{eavtrip431}
E^{(+)}_2 &= -54, & E^{(+)}_8 &= -2(20+\alpha_+), \nonumber\\
E^{(+)}_4 &= -50, & E^{(+)}_6 &= -2(20+\alpha_-), \nonumber \\
E^{(+)}_1 &= -38, & E^{(+)}_5 &= -2(20-\alpha_-), \nonumber\\
E^{(+)}_3 &= -30, & E^{(+)}_7 &= -2(20-\alpha_+),
\end{align}
with $\alpha_\pm=\sqrt{25\pm12\sqrt2}$. The integer eigenvalues belong to
mirror-even eigenvectors and the irrational ones to mirror-odd
eigenvectors, alternating as stated above. Every frequency $E_k-E_l$ is an integer combination of $1$, $\alpha_+$ and
$\alpha_-$, so $K_I=2$; the explicit form of $P^{(+)}(\tau)$ is given in
Appendix~\ref{app:ptexplicit}.

For the singlet we take $(v,w,r)=(8,9,6)$, for which
$\sqrt{v^2+r^2}=10$, $\sqrt{v^2+(w+r)^2}=17$ and
$\sqrt{2(v^2+2r^2)}=4\sqrt{17}$, giving
\begin{align}
\label{eavsing896}
E^{(-)}_2 &= -134, & E^{(-)}_8 &= -2(48+\gamma_+), \nonumber\\
E^{(-)}_4 &= -130, & E^{(-)}_6 &= -2(48+\gamma_-),\nonumber\\
E^{(-)}_1 &= -94,  & E^{(-)}_5 &= -2(48-\gamma_-),\nonumber\\
E^{(-)}_3 &= -62,  & E^{(-)}_7 &= -2(48-\gamma_+),
\end{align}
with $\gamma_\pm=\sqrt{181\pm36\sqrt{17}}$. Again the integer eigenvalues
are mirror-even, the irrational ones mirror-odd, and $K_I=2$.

\subsection{Perfect transfer}
\label{sec:pt}

Neither sector exhibits PT for any positive couplings. By
Eq.~\eqref{ephase}, PT at a time $\tau_{\rm PT}$ requires that eigenvalues of equal
parity differ by integer multiples of $2\pi/\tau_{\rm PT}$, and that eigenvalues of
opposite parity differ by odd multiples of $\pi/\tau_{\rm PT}$. In the triplet sector, $E_1^{(+)},\dots,E_4^{(+)}$ are
mirror-even and $E_5^{(+)},\dots,E_8^{(+)}$ are mirror-odd. Applying the
first condition within each parity class requires $v$, $w$,
$\sqrt{v^2+w^2}$ and $X_\pm$ all to be integer multiples of $\pi/(2\tau_{\rm PT})$.
Since $v,w>0$, the identity $X_+^2-X_-^2=2\sqrt2\,vw$ then implies
$\sqrt2\in\mathbb{Q}$, a contradiction; the triplet sector therefore
admits no PT.

In the singlet sector, with $A\equiv\sqrt{v^2+r^2}$,
$B\equiv\sqrt{v^2+(w+r)^2}$ and $Y_\pm$ the nested roots of
Eq.~\eqref{eenersing}, the same conditions make $A$, $B$, $w$ and $Y_\pm$
integer multiples of $\pi/(2\tau_{\rm PT})$, and the identity
$Y_+^2+Y_-^2=2(A^2+w^2)$ is then incompatible, modulo~4, with the parity
conditions, so this sector admits no PT either.
We therefore turn to PGT.

\subsection{Pretty good transfer}\label{sec:pgt}

\begin{figure}[tbp]
    \centering
    \includegraphics[width=1\linewidth]{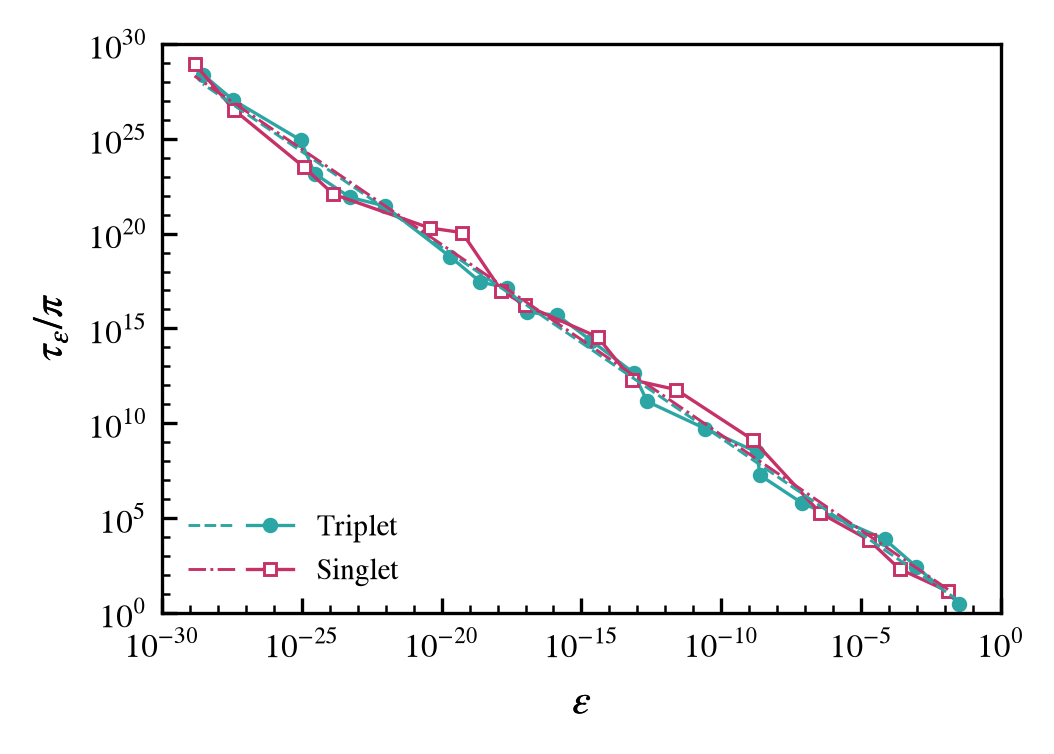}
    \caption{Times $\tau_\epsilon/\pi$ at which $P(\tau_\epsilon)=1-\epsilon$, for the
    triplet (teal, $(v,w)=(4,3)$) and the singlet (rose, $(v,w,r)=(8,9,6)$).
    Dashed lines are linear fits in log-log scale, with fitted exponents 1.008
    (teal) and 1.005 (rose).}
    \label{fig:pgt_scaling}
\end{figure}

In both examples $K_I=2$, so Eq.~\eqref{este} predicts
$\tau_\epsilon\sim\epsilon^{-1}$.
Figure~\ref{fig:pgt_scaling} tests this for the triplet transfer
$|1_+\rangle\to|8_+\rangle$ with $(v,w)=(4,3)$ and the singlet transfer
$|1_-\rangle\to|8_-\rangle$ with $(v,w,r)=(8,9,6)$. It shows, on a log-log
scale, times $\tau_\epsilon$ at which $P(\tau_\epsilon)=1-\epsilon$, obtained as
described below; each is an upper bound on the arrival time for its
$\epsilon$. Linear fits give exponents 1.008 (triplet) and 1.005
(singlet), in agreement with $K_I/2=1$.

\begin{figure}[tbp]
    \centering
    \includegraphics[width=1\linewidth]{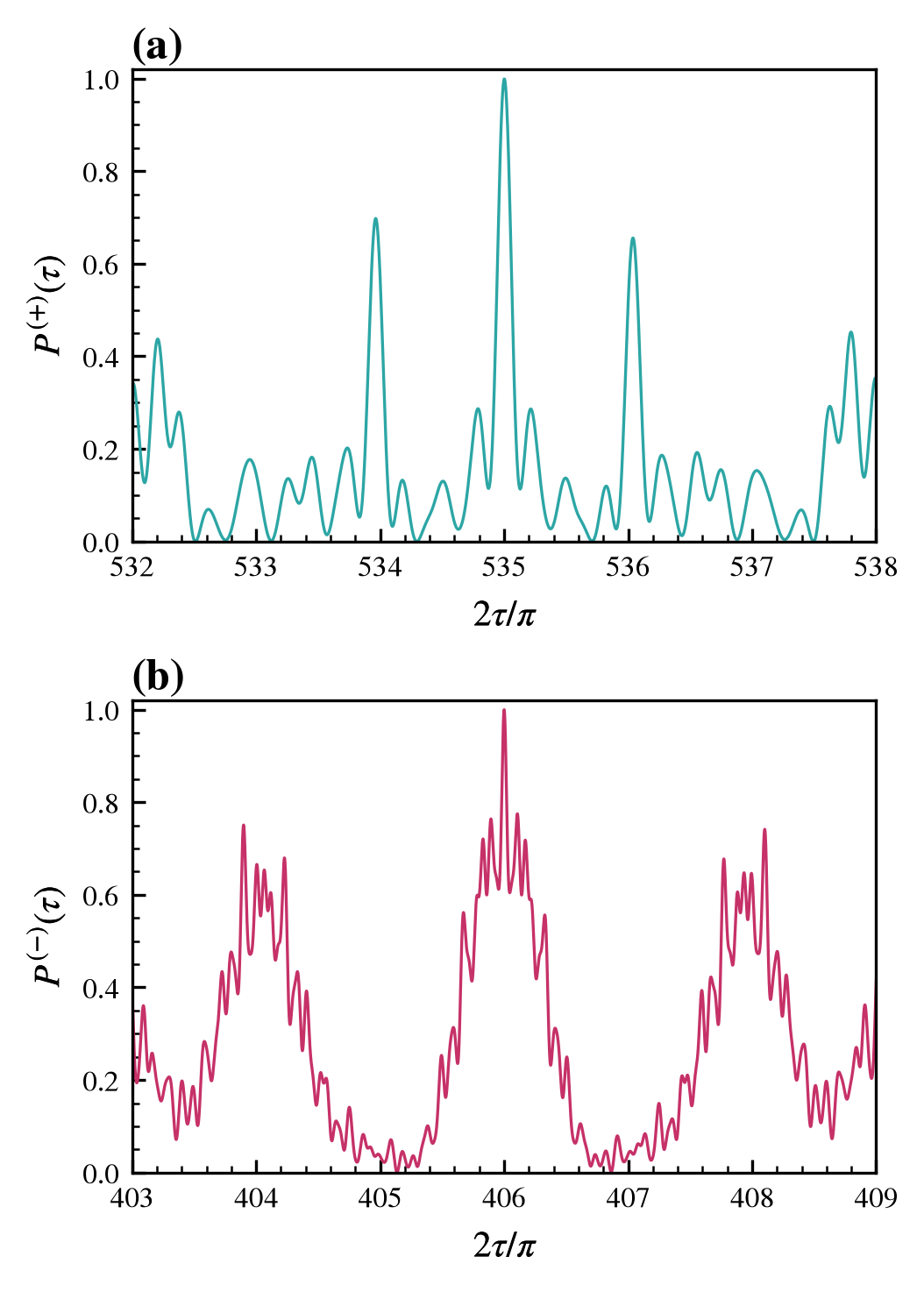}
    \caption{$P^{(+)}(\tau)$ for $v=4$, $ w=3$, and $r=1$ in the neighborhood of $\tau_2$ of Table~\ref{ttrip}, and $P^{(-)}(\tau)$ for $v=8, w=9$ and $r=6$ in the neighborhood of $\tau_2$ of Table~\ref{tsing}. }
    \label{fig:p_sing_896_trip_431}
\end{figure}

Figure~\ref{fig:pgt_scaling} was obtained as follows. At times $\tau=q\pi/2$
with integer $q$, the integer eigenvalues in Eqs.~\eqref{eavtrip431} and
\eqref{eavsing896} are all twice an odd number, so the mirror-even
eigenvectors acquire the common phase $(-1)^q$ (other values of $r$ only
add a global phase). The mirror-odd eigenvectors acquire
$e^{\pm i\pi q\alpha_\pm}$ [or $e^{\pm i\pi q\gamma_\pm}$]. By
Eq.~\eqref{ephase}, $P(q\pi/2)$ is close to one when
$e^{\pm i\pi q\alpha_\pm}\approx(-1)^{q+1}$, i.e.\ when $q\alpha_+$ and
$q\alpha_-$ are both close to integers with the parity of $q+1$.
Continued fractions give the best rational approximations of a single
number, but there is no equally simple algorithm for approximating two
numbers with a common denominator $q$, so we find such $q$ with the
Lenstra--Lenstra--Lov\'asz (LLL) lattice basis reduction
algorithm~\cite{lagariasComputationalComplexitySimultaneous1985,bosmaFindingSimultaneousDiophantine2013a}. We denote the successive values,
ordered by increasing accuracy, by $q_j$, and the corresponding times by
$\tau_j=q_j\pi/2$. They are listed in Tables~\ref{ttrip} and \ref{tsing}.
Figure~\ref{fig:p_sing_896_trip_431} shows $P(\tau)$ around $\tau_2$: the
high-fidelity peak is isolated and sharp.

\begin{table}[tbp]
\begin{center}
\begin{tabular}{c|l|l}
 $j$& $q_j$&   $P^{(+)}(\tau_j=q_j\,\pi/2)$  \\ \hline
  1&  6  & 0.97 \\
 2 &  535   & 0.9990\\
3  & 14816 & 0.99993 \\
4 & 1216279 & 0.99999992\\
5 & 36144584 & 0.999999998\\
6&  10440045472&  0.99999999997  \\
7&60411866963 & 0.999999999990\\
8 & 278986982586&  0.9999999999998 \\
9 & 8958903579168 &0.99999999999993 \\
10&446431612572829 &  0.999999999999998\\
\end{tabular}
\end{center}
\caption{Values of $q_j$ and $P^{(+)}(\tau_j)$ for $v=4$ and $ w=3$  using
rational approximations
for the irrational numbers appearing in the eigenvalues 
Eq.~\eqref{eavtrip431}: $\sqrt{25 \pm12 \sqrt{2}}$.
 \label{ttrip}}
\end{table}

\begin{table}[tbp]
\begin{center}
\begin{tabular}{c|l|l}
 $j$& $q_j$&   $P^{(-)}(\tau_j=q_j\,\pi/2)$  \\ \hline
  1&27 &0.990  \\
 2 & 406& 0.9998\\
3  &13843 &0.99998 \\
4 & 360757 &0.9999996 \\
5 &108543653 &0.999999990 \\
6 & 2560056723& 0.9999999990 \\
7 & 1177764129958& 0.999999999997 \\
8 &  3842800293927 &0.99999999999993\\
9 & 702486791398912& 0.999999999999996 \\
10&33937849905939100 & 0.999999999999999990 \\
\end{tabular}
\end{center}
\caption{Values of $q_j$ and $P^{(-)}(\tau_j)$ for  $v=8, w=9$ and $r=6$ using
rational approximations
for the irrational numbers appearing in the eigenvalues 
Eq.~\eqref{eavsing896}:
 $\sqrt{181 \pm 36\sqrt{17}}$.
 \label{tsing}}
\end{table}

We now ask whether PGT also holds between bulk sites. The condition for
high transfer at $\tau_j$, Eq.~\eqref{ephase}, does not depend on the site.
The same times $\tau_j$ therefore give high transfer between every mirror
pair of the triplet chain, $|j_+\rangle\to|(9-j)_+\rangle$. Applying the
transfer twice returns the state to its origin, so the retrieval
probability $P^{(+)}_{1\to1}$ is also close to one at $2\tau_j=q_j\pi$.
Table~\ref{ttripb} confirms this for $(v,w)=(4,3)$: all these
probabilities approach one as $j$ increases.
 
At the times $\tau=q\pi/2$ we also find $P^{(+)}_{3\to6}=P^{(+)}_{2\to7}$ and
$P^{(+)}_{4\to5}=P^{(+)}_{1\to8}$, although these probabilities differ at
generic times. This follows from a symmetry of the mirror-odd eigenvectors
that relates sites $i$ and $5-i$. Table~\ref{ttripb} therefore lists only
$P^{(+)}_{2\to7}$ and $P^{(+)}_{4\to5}$.
 
\begin{table*}
% \begin{ruledtabular}
\begin{tabular}{c|l|l|l|l}
 $j$& $q_j$&   $P^{(+)}_{2 \to 7}(\tau_j=q_j\,\pi/2)$ & $P^{(+)}_{4 \to 5} (\tau_j=q_j\,\pi/2)$& $P^{(+)}_{1 \to 1} (\tau_j=q_j\,\pi)$ \\ \hline
  1&  6  & 0.962& 0.968 &0.88\\
 2 &  535   &0.990 & 0.9990 & 0.996 \\
3  & 14816 &  0.99995 & 0.99993 & 0.9997 \\
4 & 1216279 &0.99999994 & 0.99999992 &0.9999997\\
5 & 36144584 & 0.999999998 & 0.999999998 & 0.999999990\\
6&  10440045472& 0.99999999998 & 0.99999999997 & 0.99999999990 \\
7&60411866963 & 0.999999999990 & 0.999999999990 & 0.99999999996 \\
8 & 278986982586& 0.9999999999998 & 0.9999999999997 & 0.9999999999991 \\
9 & 8958903579168 & 0.99999999999995 &0.99999999999993 & 0.9999999999997\\
10&446431612572829 & 0.999999999999998 & 0.999999999999997 & 0.999999999999991 \\
\end{tabular}
\caption{Bulk transfer and retrieval in the triplet sector for $(v,w)=(4,3)$,
at the times $\tau_j=q_j\pi/2$ of Table~\ref{ttrip} (retrieval at
$2\tau_j=q_j\pi$). At these times $P^{(+)}_{3\to6}=P^{(+)}_{2\to7}$, and
$P^{(+)}_{4\to5}$ coincides with $P^{(+)}_{1\to8}$ in Table~\ref{ttrip}.
\label{ttripb}}
\end{table*}

\subsection{Almost perfect transfer}\label{sec:apt}
The scaling law of Eq.~\eqref{este} makes PGT alone impractical, but a
network that exhibits PGT is a natural candidate for APT at experimentally
relevant times. Since rescaling all couplings by $\lambda$ rescales time by
$1/\lambda$ (Sec.~\ref{subsec:heisenberg_ham}), any target time could be met by
speeding up the dynamics, so APT must be defined at a fixed energy scale. We
keep $w$ fixed and say that the network exhibits APT for a tolerance $\Delta$
at a time $\tau_\Delta$ if
\begin{equation}
\label{apst}
\exists \; (v_\Delta, r_\Delta) :\quad P(\tau_\Delta)>1-\Delta .
\end{equation}
Here $\Delta$ is the maximum tolerable transmission error and $\tau_\Delta$
an experimentally realistic time.
Reference~\cite{vinetAlmostPerfectState2012} uses the name almost perfect state transfer for the property called PGT here (see also Ref.~\cite{banchiPrettyGoodState2017a}); we reserve the name APT for the finite-time criterion of Eq.~\eqref{apst}.
 
\begin{figure}[tbp]
    \centering
    \includegraphics[width=1\linewidth]{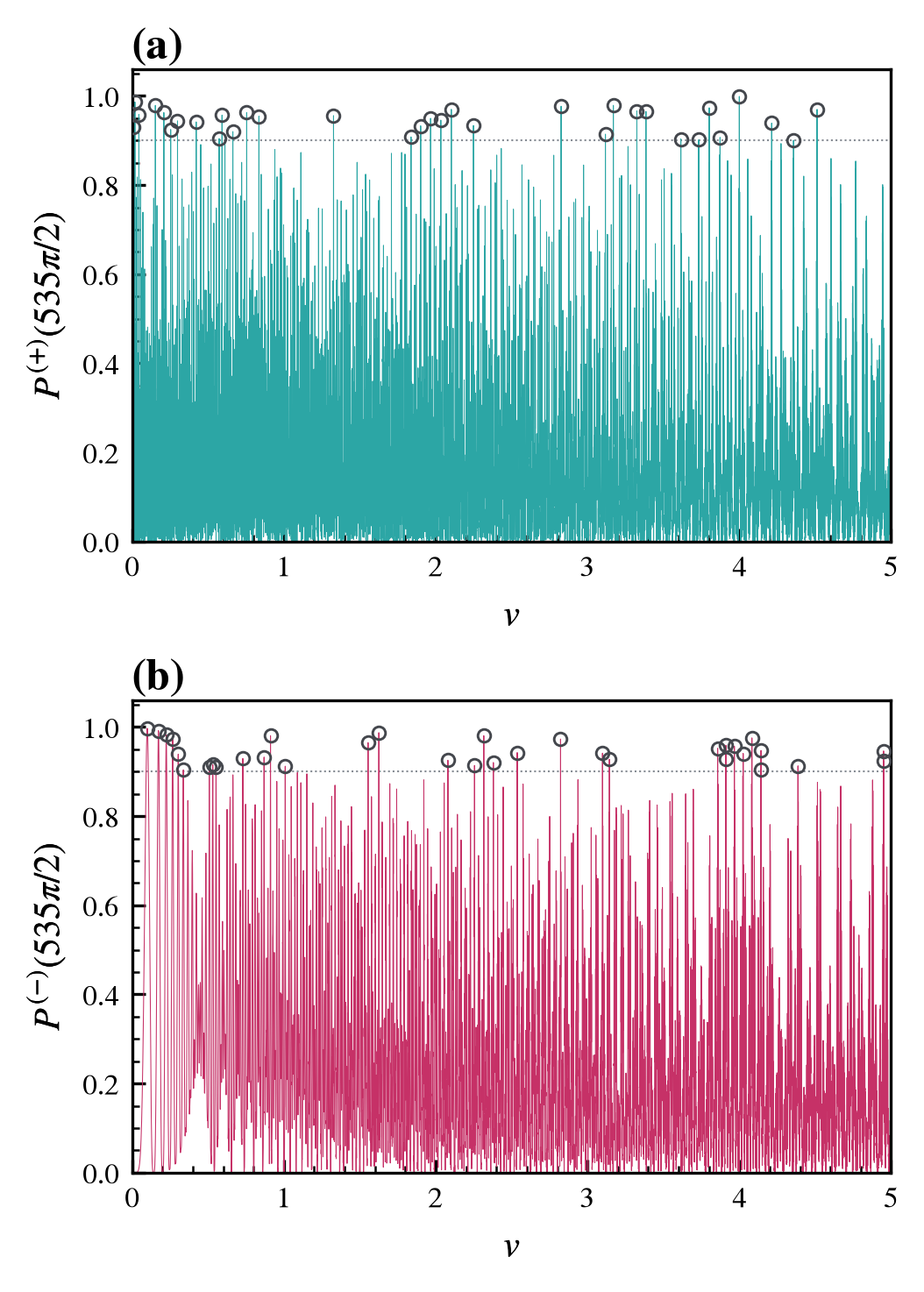}
    \caption{(a) shows $P^{(+)}(\tau=535 \pi/2)$ as a function of $v$ for $w=3$ in the triplet state. The dark gray circles show the 33 peaks where $P^{(+)}>0.9$, corresponding to $\Delta<0.1$.  \\ (b) shows $P^{(-)}(\tau=535 \pi/2)$ as a function of $v$ for $w=3, r=1$ in the singlet  state. The dark gray circles show the 34 peaks where $P^{(-)}>0.9$, corresponding to $\Delta<0.1$.}
    \label{fig:ptrip_sing_w3_t2}
\end{figure}

As an example we keep $w=3$ and fix $\tau=535\pi/2$, the time $\tau_2$ of
Table~\ref{ttrip} at which $(v,w)=(4,3)$ already gives
$P^{(+)}=0.9990$, and look for high values of the transfer probability as
a function of $v$, sampled in steps of $10^{-4}$ for the triplet and
$2\times10^{-4}$ for the singlet. The probabilities
$P^{(+)}(\tau=535\pi/2)$ and $P^{(-)}(\tau=535\pi/2)$, the latter with $r=1$,
are shown in Fig.~\ref{fig:ptrip_sing_w3_t2}.

\subsubsection*{Tuning in $P^{(-)}$}
 
The probabilities are highly oscillating functions, and for a given time at which $P^{(+)}$ is almost zero or one, $P^{(-)}$  could take any value between zero and one. The fact that $P^{(+)}$ does not depend on $r$ may be used to tune  $P^{(-)}$, either close to one, or close to zero. Examples of the four possible situations are given in Table~\ref{ttps}.
 
\begin{table}[htbp]
\begin{center}
\begin{tabular}{c|c|l|l}
$ \tau$& $r$&   $P^{(+)}_{1 \to 8}(\tau)$ & $P^{(-)}_{1 \to 8} (\tau)$ \\ \hline
$ 535 \pi/2$ &  9.8889  & 0.9990& 0.997 \\
$ 535 \pi/2$  &4.828 & 0.9990 & 0.00003 \\
$ 535 \pi$ &  6.111 & 0.0007 & 0.993 \\
$ 535 \pi$ & 5.929 & 0.0007 &0.000007\\
\end{tabular}
\end{center}
\caption{Values of $P^{(+)}_{1 \to 8}(\tau)$ and $P^{(-)}_{1 \to 8}(\tau)$ 
close to one or close to zero for $v=4, w=3$ and selected values of $\tau$ and 
$r$.
 \label{ttps}}
\end{table}
 
Then, suppose we have to transmit a triplet state, but the preparation leaves a small singlet component,
 
\begin{equation}
\label{epsi0}
|\psi(0)\rangle = \omega_+ |1_+\rangle + \omega_- |1_-\rangle ,
\end{equation}
where $|\omega_+|^2+ |\omega_-|^2=1$ and $ |\omega_+|^2\gg |\omega_-|^2$.

Since $H_N$ does not couple
the two sectors, the probabilities of finding the triplet and the singlet at
the target are
\begin{equation}
\bigl|\langle 8_\pm|e^{-iH_N \tau}|\psi(0)\rangle\bigr|^2=|\omega_\pm|^2\,P^{(\pm)}(\tau).
\label{eq:Ppsi}
\end{equation}

To suppress the singlet component for $(v,w)=(4,3)$, we can set $r=4.828$.
At $\tau=535\pi/2$, Table~\ref{ttps} then gives a singlet probability
$|\omega_-|^2P^{(-)}=3\times10^{-5}\,|\omega_-|^2$, the product of two small
quantities, while the triplet is still transferred with $P^{(+)}=0.9990$.

% ########################################

\subsection{Robustness to coupling disorder}
\label{sec:disorder}
 
In an experiment the couplings will not take exactly their design values, and
static disorder is known to degrade state transfer in spin
chains~\cite{dechiaraPerfectFractalTransmission2005,petrosyanStateTransferStatic2010,zwickRobustnessSpincouplingDistributions2011}.
We model it by replacing each coupling $J_b\in\{v,w,r\}$ of
Eq.~\eqref{eq:HN} by $J_b(1+u_b)$, with $u_b$ drawn uniformly from
$[-\eta,\eta]$ independently for every bond, and average over 400
realizations. We compare the triplet transfer $|1_+\rangle\to|8_+\rangle$
with the end-to-end transfer of an eight-site Heisenberg SSH chain with the
same couplings $v$ and $w$. Without disorder the two are identical, since the
triplet block is the single-arm block shifted by a constant
[Appendix~\ref{app:rindep} and Ref.~\cite{serraExactSolutionFamily2022}], so
any difference comes from how the two structures respond to disorder. Because
disorder shifts the transfer peaks by a different amount in each realization,
we measure the best transfer within $\tau_0\pm0.5$, where $\tau_0$ is the time of a
transfer peak of the clean system. Unless stated otherwise, $\eta=0.05$,
$(v,w)=(4,3)$ and $\tau_0=3\pi$, the time $\tau_1$ of Table~\ref{ttrip}, where the
best clean transfer in this window is $0.976$.
 
\begin{figure}[tbp]
    \centering
    \includegraphics[width=1\linewidth]{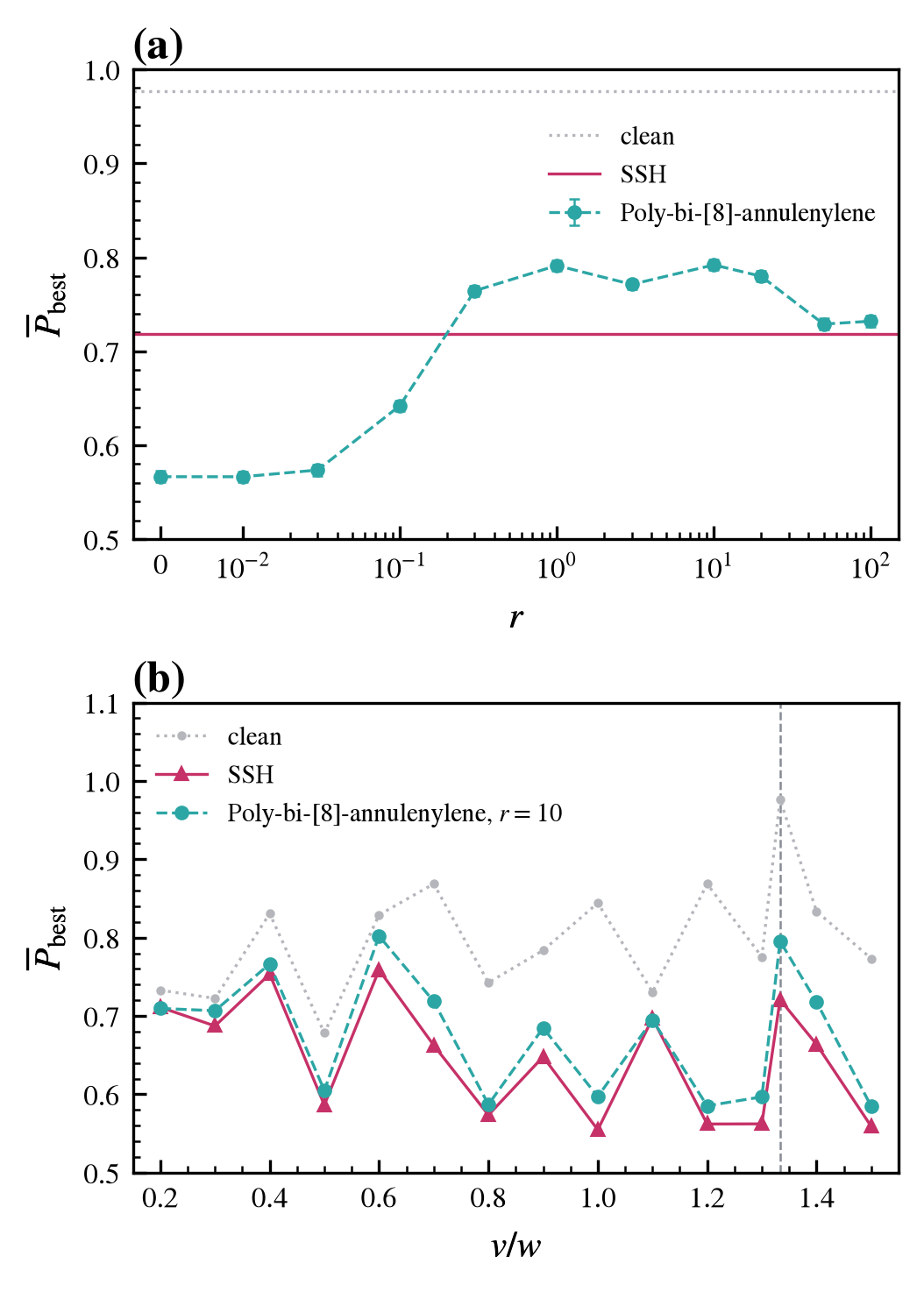}
    \caption{Triplet transfer under coupling disorder of strength
    $\eta=0.05$: best transfer within $\tau_0\pm0.5$ of the clean peak time
    $\tau_0$, averaged over 400 realizations.
    (a)~As a function of $r$ for $(v,w)=(4,3)$ and $\tau_0=3\pi$; $r=0$ is
    plotted at the left edge. Error bars give the standard error of the mean. The horizontal lines give the clean value and
    the eight-site Heisenberg SSH chain. (b)~As a function of $v/w$ for $w=3$
    and $r=10$, with $\tau_0$ the highest clean peak in $0<\tau<30$; the dashed line
    marks $v/w=4/3$.}
    \label{fig:disorder}
\end{figure}

Figure~\ref{fig:disorder}(a) shows the result as a function of $r$. The SSH
chain reaches $0.72$. Without rungs the lattice does worse, $0.57$: the two
components of $|1_+\rangle$ travel along two arms with different disorder and
lose their relative phase. For $r\gtrsim0.2$ the lattice overtakes the SSH
chain, and it reaches $0.77$--$0.79$ for $1\lesssim r\lesssim20$. The rungs
lock the two arms together, so the triplet sees the average of the disorder on
the two arms, which fluctuates less than the disorder on either arm.
Consistently, the advantage disappears when the disorder is the same on
mirror-image bonds ($0.72$ at $r=10$), and disorder on the rungs themselves
does not affect the triplet transfer, because the rungs only shift the triplet
block by a constant [Eq.~\eqref{ehtr}]. Very strong rungs help less ($0.73$ at
$r=50$ and $100$), so the rung coupling has an optimal range.
 
Figure~\ref{fig:disorder}(b) repeats the comparison for other coupling ratios
at $r=10$. The lattice is never significantly below the SSH chain. The gain
depends on the coupling ratio: it is negligible at $v/w=0.2$ and $1.1$, where
disorder affects the peak only weakly, and largest, $0.07$, at $(v,w)=(4,3)$.
 
The late peak at $\tau=535\pi/2$ used in Secs.~\ref{sec:pgt} and~\ref{sec:apt}
is much more fragile, because the errors in the phases grow with time. At
$\eta=10^{-3}$ the best triplet transfer near this time falls from
$0.9994$ to $0.69$, and the singlet transfer, suppressed to $3\times10^{-5}$
with $r=4.828$ in Table~\ref{ttps}, rises to $0.09$; only
$\eta\approx10^{-4}$ preserves both ($0.992$ and $0.001$). For realistic
disorder the early peaks are therefore the useful ones, as we also find for
the Hubbard model below.

% ########################################

\section{Transfer in the Hubbard Model}\label{sec:hubbard}

The Heisenberg couplings of Eq.~\eqref{eq:HN} arise naturally as the superexchange of
electrons, or of fermionic atoms in an optical lattice, in a half-filled Hubbard model.
Unlike the spin model, the Hubbard model retains charge fluctuations: an electron can
virtually or really occupy a site together with a second electron. State transfer has also been
studied for particles hopping on engineered bosonic and fermionic
networks~\cite{yungPerfectStateTransfer2005}; the Hubbard model adds the on-site interaction that
generates the exchange couplings. Here we ask which
features of the transfer found above survive in this setting. The analysis is exact
within the relevant sector of the Hubbard model and involves no truncation.

\subsection{Model and strong-coupling limit}\label{sec:hubbard-model}

On the network of Fig.~\ref{fig:pb8a_N2} we consider
\begin{equation}
  H_{\rm Hub} = -\sum_{b=\langle ij\rangle} t_b \sum_{\sigma=\uparrow,\downarrow}
  \big(c^\dagger_{i\sigma}c_{j\sigma}+c^\dagger_{j\sigma}c_{i\sigma}\big)
  + U\sum_i n_{i\uparrow}n_{i\downarrow},
  \label{eq:hubbard}
\end{equation}
with hopping $t_b\in\{t_v,t_w,t_r\}$ on the bonds labeled $v$, $w$, $r$, and on-site
repulsion $U>0$. At half filling and $U\gg|t_b|$, degenerate perturbation theory
(equivalently a Schrieffer--Wolff transformation~\cite{schrieffer1966, bravyi2011}) reduces Eq.~\eqref{eq:hubbard} to the
antiferromagnetic Heisenberg model~\cite{chao1977,auerbach1994,fazekas1999}
\begin{equation}\label{eq:h_eff}
    H_{\rm eff}=\sum_b (4t_b^2/U)\,(\bm{S}_i\cdot\bm{S}_j-\tfrac14)+O(t_b^4/U^3).
\end{equation}
With $\bm{S}_i=\bm{\sigma}_i/2$, in the Pauli normalization of Eq.~\eqref{eq:HN} each bond contributes
$(t_b^2/U)(\bm{\sigma}_i\cdot\bm{\sigma}_j-1)$, so the couplings are
\begin{equation}
  v=\frac{t_v^2}{U},\qquad w=\frac{t_w^2}{U},\qquad r=\frac{t_r^2}{U}.
  \label{eq:parameter-map}
\end{equation}
We take the strong-coupling limit $U/t_b\to\infty$ at fixed $(v,w,r)$, i.e.\ with
$t_b=\sqrt{J_bU}$ for $J_b\in\{v,w,r\}$. The exchange couplings, and with them every time
scale of the spin dynamics, then stay fixed, so the Hubbard and Heisenberg dynamics can be
compared at the same time $\tau$ ($\hbar=1$, energies in the units of Eq.~\eqref{eq:HN}). We quantify the interaction strength by $U/t_v$.

The couplings obtained this way are antiferromagnetic, whereas Eq.~\eqref{eq:HN} is
ferromagnetic. Writing $H_{\rm FM}\equiv-\sum_b J_b\,\bm{\sigma}_i\cdot\bm{\sigma}_j$
for Eq.~\eqref{eq:HN}, the leading term of Eq.~\eqref{eq:h_eff} is
\begin{equation}\label{eq:fm-afm}
  \begin{aligned}
    H_{\rm AFM}&\equiv\sum_b J_b\big(\bm{\sigma}_i\cdot\bm{\sigma}_j-1\big)
    =-H_{\rm FM}-J_{\rm tot},\\
    J_{\rm tot}&\equiv\sum_b J_b=8v+6w+4r,
  \end{aligned}
\end{equation}
where $J_{\rm tot}$ sums the couplings over the 8 $v$, 6 $w$ and 4 $r$ bonds of the network.
Every eigenstate $\ket{\phi_k}$ of $H_{\rm FM}$ with energy $E_k$ is therefore an eigenstate of $H_{\rm AFM}$
with energy $-E_k-J_{\rm tot}$: the spectrum is reflected and rigidly shifted. The shift fixes
the zero of energy. The polarized state $\ket{F_\uparrow}$, with $H_{\rm FM}\ket{F_\uparrow}=-J_{\rm tot}\ket{F_\uparrow}$,
has $H_{\rm AFM}$ energy $0$, which is also its exact Hubbard energy, since Pauli blocking forbids every hop in the fully polarized half-filled state. It is therefore the highest state of $H_{\rm AFM}$, not the ground state; the transfer protocol requires only that it be an exact eigenstate.

The sign does not affect the transfer.
By Eq.~\eqref{eamp}, $P(\tau)=\sum_{k,l}s_ks_l\,p_kp_l\cos[(E_k-E_l)\tau]$, with the
parities $s_k$ and weights $p_k$ defined there, where $k$ and $l$ both run over the
eigenvectors $\ket{\phi_k}$ of $H_{\rm FM}$ in the chosen sector. By Eq.~\eqref{eq:fm-afm}, $E_k-E_l\to-(E_k-E_l)$: the constant $J_{\rm tot}$
cancels from every difference, and the sign reversal leaves each cosine unchanged.
A ferromagnetic realization is also available with two-component bosons~\cite{duan2003}. Their superexchange has been observed directly~\cite{trotzky2008} and
used to follow a single spin impurity moving on a polarized background~\cite{fukuhara2013}.

\subsection{The transfer problem in the Hubbard model}\label{sec:hubbard-transfer}

\emph{Sector.} We take the single excitation to be one down spin on the polarized state
$\ket{F_\uparrow}$, i.e.\ $N_\uparrow=15$ and $N_\downarrow=1$, which Eq.~\eqref{eq:hubbard}
conserves separately. Flipping all spins, which commutes with Eqs.~\eqref{eq:HN}
and~\eqref{eq:hubbard}, maps this onto the single up spin of Sec.~\ref{subsec:heisenberg_ham},
so the transfer probabilities are the same in both conventions. The sector is spanned by the 256 states
\begin{equation}
  \ket{d,h}=c^\dagger_{d\downarrow}c_{h\uparrow}\ket{F_\uparrow},\qquad
  \ket{F_\uparrow}=c^\dagger_{1\uparrow}c^\dagger_{2\uparrow}\cdots c^\dagger_{16\uparrow}\ket{0},
\end{equation}
labeled by the position $d$ of the down electron and the site $h$ from which an up electron is
missing. For $d=h$ this is the single spin flip $\ket{n,n}=S^-_n\ket{F_\uparrow}$; for $d\neq h$
site $d$ is doubly occupied and site $h$ empty. Equation~\eqref{eq:hubbard} acts as
\begin{multline}
  H_{\rm Hub}\ket{d,h}=U(1-\delta_{dh})\ket{d,h}\\
  -\sum_{d'}t_{dd'}\ket{d',h}+\sum_{h'}t_{hh'}\ket{d,h'},
  \label{eq:two-body}
\end{multline}
with $t_{ij}$ the hopping on bond $(i,j)$. The sector is therefore \emph{exactly} a two-body
problem: a down particle hopping with $-t_b$ and an up hole hopping with $+t_b$, with an on-site
attraction $-U$ relative to the continuum. The single spin flips (magnons) are its bound states.
For $U\gg t_b$ there are 16 bound states near $E=0$ and 240 doublon--holon states near $E=U$.
The mirror symmetries of Fig.~\ref{fig:pb8a_N2} are even permutations of the sites, so they
leave $\ket{F_\uparrow}$ invariant and act on $\ket{d,h}$ without fermionic signs. The two sectors
of opposite chain-exchange parity (128 states each) therefore remain exactly decoupled at any
$U$.

\emph{Initial and target states.} The pair states of Eq.~\eqref{eq:pair-states} become
\begin{align}
  \ket{1_\pm}&=\tfrac{1}{\sqrt2}\big(S^-_1\pm S^-_{16}\big)\ket{F_\uparrow}=\tfrac{1}{\sqrt2}\big(\ket{1,1}\pm\ket{16,16}\big),\nonumber\\
  \ket{8_\pm}&=\tfrac{1}{\sqrt2}\big(\ket{8,8}\pm\ket{9,9}\big).
  \label{eq:hubbard-pair-states}
\end{align}
Because $S^-_n=c^\dagger_{n\downarrow}c_{n\uparrow}$ is a bilinear, these carry no fermionic sign
ambiguity: $\ket{1_+}$ and $\ket{1_-}$ are the triplet and singlet states of the pair $(1,16)$, as in
the spin model.

\begin{figure*}[!htb]
  \includegraphics[width=\textwidth]{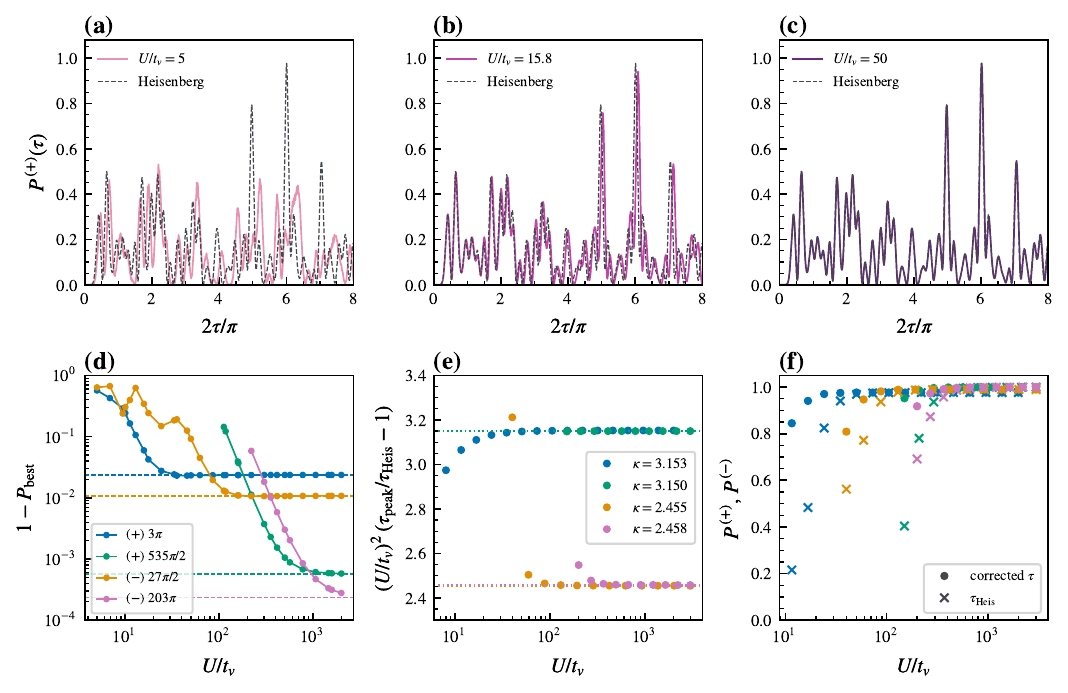}
  \caption{Transfer in the Hubbard model. (a)--(c) $P^{(+)}(\tau)$ for $(v,w,r)=(4,3,1)$ at
  $U/t_v=5$, $15.8$ and $50$ (solid), compared with the Heisenberg model (dashed). (d) Infidelity, $1-P_{\rm best}$, in the windows $3\pi\pm0.5$ and $535\pi/2\pm3$ [triplet (+)] and
  $27\pi/2\pm0.5$ and $203\pi\pm3$ [singlet ($-$), $(8,9,6)$]; dashed lines give the Heisenberg values
  in the same windows. Below $U/t_v\approx100$ ($535\pi/2$) and $200$ ($203\pi$) a neighboring peak in
  the window is higher than the delayed one; these points are omitted.
  (e) Relative delay of the transfer peak (peak of the bound-band amplitude) in units of $(t_v/U)^2$. Dotted lines give its limit $\kappa$ of
Eq.~\eqref{eq:delay}; colors as in (d).
  (f) Transfer at the Heisenberg peak time
  $\tau_{\rm Heis}$ ($\times$) and at the corrected time $\tau_{\rm Heis}[1+\kappa\,(t_v/U)^2]$ ($\bullet$); colors as in (d).}
  \label{fig:hub-transfer}
\end{figure*}

\emph{Transfer amplitude.} The triplet and singlet transfer probabilities are defined as in
Eq.~\eqref{eP}, $P^{(\pm)}(\tau)=|\langle 8_\pm|e^{-iH_{\rm Hub}\tau}|1_\pm\rangle|^2$, and the
amplitude has the form of Eq.~\eqref{eamp},
\begin{equation}
\langle 8_\pm|e^{-iH_{\rm Hub}\tau}|1_\pm\rangle=\sum_{k=1}^{128}s_k\,p_k\,e^{-iE_k\tau},
\label{eq:hub-amp}
\end{equation}
where $\ket{\phi_k}$ and $E_k$ are now the eigenvectors and energies of $H_{\rm Hub}$ in the
chosen sector, taken with definite parity $s_k$ under the reflection that exchanges the ends
(Sec.~\ref{sec:transfer}), and $p_k=|\langle 1_\pm|\phi_k\rangle|^2$. Keeping only the eight
bound states in the sum gives the bound-band amplitude of Fig.~\ref{fig:hub-transfer}(e).

\emph{Observable.} In this sector $n_{n\downarrow}(1-n_{n\uparrow})=\ket{n,n}\bra{n,n}$. Hence,
exactly at any $U$,
\[
\begin{aligned}
P^{(\pm)}(\tau)
&=\big|\langle 8_\pm|e^{-iH_{\rm Hub}\tau}|1_\pm\rangle\big|^2\\
&=2\left|\langle 8,8|e^{-iH_{\rm Hub}\tau}|1_\pm\rangle\right|^2
=2\big\langle n_{8\downarrow}(1-n_{8\uparrow})\big\rangle_\tau ,
\end{aligned}
\]
where $\langle\cdot\rangle_\tau$ is the expectation value in $e^{-iH_{\rm Hub}\tau}\ket{1_\pm}$.
The second line uses Eq.~\eqref{eq:hubbard-pair-states} and chain-exchange symmetry,
$\langle 9,9|e^{-iH_{\rm Hub}\tau}|1_\pm\rangle=\pm\langle 8,8|e^{-iH_{\rm Hub}\tau}|1_\pm\rangle$.
The transfer probability is thus a single-site, spin- and density-resolved occupation, as
measured in quantum-gas microscopes~\cite{boll2016,parsons2016}.

All results below are obtained by exact diagonalization of the $256\times256$ matrix of
Eq.~\eqref{eq:two-body}. We use the triplet example $(v,w,r)=(4,3,1)$ of
Secs.~\ref{sec:pgt} and \ref{sec:apt} and the singlet example $(8,9,6)$ of Sec.~\ref{sec:pgt}.

\subsection{Spectrum}\label{sec:hubbard-spectrum}

\begin{figure}[!tb]
  \includegraphics[width=\columnwidth]{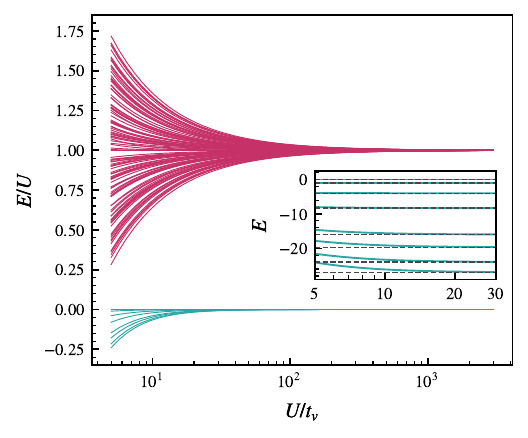}
  \caption{Spectrum of the Hubbard model in the triplet (+) part of the single-excitation sector
  $(N_\uparrow,N_\downarrow)=(15,1)$, for $(v,w,r)=(4,3,1)$ with $t_b=\sqrt{J_bU}$.
  Main panel: all 128 levels in units of $U$, with eight bound, magnon-like states near zero
  (teal) and 120 doublon--holon states near $U$ (rose). Inset: the bound band (solid) for
  $5\le U/t_v\le30$ and the Heisenberg prediction
  $E=-E^{(+)}_k-J_{\rm tot}$ (dashed), with $E^{(+)}_k$ from Eq.~\eqref{eenertri} and
  $J_{\rm tot}$ from Eq.~\eqref{eq:fm-afm}.}
  \label{fig:hub-spectrum}
\end{figure}

Figure~\ref{fig:hub-spectrum} shows the structure implied by Eq.~\eqref{eq:two-body}. For
$U/t_v\gtrsim5$ a band of eight bound states in each parity sector is separated from the
doublon--holon continuum, and the bound states carry most of their weight on $d=h$. The bound
energies converge to the Heisenberg spectrum reflected and shifted as in Eq.~\eqref{eq:fm-afm},
$E=-E^{(+)}_k-J_{\rm tot}$, with $E^{(+)}_k$ the triplet energies of Eq.~\eqref{eenertri}
[inset of Fig.~\ref{fig:hub-spectrum}]. The level ordering is inverted: the uniform magnon, the
ferromagnetic ground state with $E^{(+)}_2=-J_{\rm tot}$, becomes the top of the band. It stays at $E=0$ exactly for every $U$, since $\sum_n\ket{n,n}$ is an exact eigenstate: applying Eq.~\eqref{eq:two-body}, the $U$ term vanishes because $d=h$, and the two hopping terms give $-\sum_{n,n'}t_{nn'}\ket{n',n}$ and $+\sum_{n,n'}t_{nn'}\ket{n,n'}$, which cancel because $t_{nn'}=t_{n'n}$.

\subsection{Approach to the Heisenberg limit and the transfer time}\label{sec:hubbard-time}

Figure~\ref{fig:hub-transfer}(a)--(c) compares the transfer probability with its Heisenberg
counterpart. At $U/t_v=5$ the charge fluctuations reshape the signal. At $U/t_v=15.8$ the peak
structure is recognizable but visibly delayed, and at $U/t_v=50$ the two curves are almost
indistinguishable. Figure~\ref{fig:hub-transfer}(d) quantifies this for the peaks studied in
Secs.~\ref{sec:pgt} and \ref{sec:apt}. Because finite $U$ shifts the peaks, we quote the best
transfer within the windows of Fig.~\ref{fig:hub-transfer}(d) in both models; for the
Heisenberg model these values slightly exceed those at exactly $\tau_j$ in
Tables~\ref{ttrip} and~\ref{tsing}:
\begin{itemize}
\item The early triplet peak near $\tau=3\pi$ reaches $0.43$, $0.76$ and $0.94$ at $U/t_v=5$, $10$
  and $15.8$, and its Heisenberg value $0.976$ from $U/t_v\approx30$ on.
\item The pretty-good-transfer peak at $\tau=535\pi/2$ requires much stronger interactions:
  $0.961$ at $U/t_v=158$ and $0.9989$ at $U/t_v=500$, against $0.9994$ in the Heisenberg model.
\item The singlet peaks follow the same pattern. The early peak near $27\pi/2$ gives $0.60$,
  $0.81$, $0.987$ and $0.989$ at $U/t_v=11.2$, $35.4$, $112$ and $354$ (Heisenberg: $0.989$),
  and the pretty-good-transfer peak at $\tau=203\pi$ gives $0.990$ and $0.997$ at
  $U/t_v=354$ and $500$ (Heisenberg: $0.9998$).
\end{itemize}
The exact $r$ independence of the triplet transfer in the Heisenberg model is lost at
$O(t_b^4/U^3)$, but the effect is small: at $U/t_v=50$ the best $3\pi$ peak varies only between
$0.9763$ and $0.9769$ for $0.5\le r\le3$ at fixed $(v,w)=(4,3)$.

The main finite-$U$ effect is a systematic delay of the high-fidelity transfer peaks,
\begin{equation}
  \tau_{\rm Hub}\simeq \tau_{\rm Heis}\left[1+\kappa\left(\frac{t_v}{U}\right)^2\right],
  \label{eq:delay}
\end{equation}
with $\kappa=3.153$ and $3.150$ for the triplet peaks at $\tau_1=3\pi$ and $\tau_2=535\pi/2$ of
Table~\ref{ttrip} and $\kappa=2.455$ and $2.458$ for the singlet peaks at $\tau_1=27\pi/2$ and
$\tau_2=203\pi$ of Table~\ref{tsing}. Here $\tau_{\rm Heis}$ is the time of the
Heisenberg maximum within the window of Fig.~\ref{fig:hub-transfer}(d), which differs slightly
from the nominal $\tau_j$ of the tables (for the first triplet peak, $\tau_{\rm Heis}=9.4366$
rather than $3\pi=9.4248$), $\tau_{\rm Hub}$ is the corresponding peak of the bound-band
amplitude, and $\kappa$ is the $U\to\infty$ limit of the scaled delay illustrated in
Fig.~\ref{fig:hub-transfer}(e).
At fixed couplings and initial state, \(\kappa\) is the same for every high-fidelity peak of a sector, up to small corrections from imperfect phase alignment, with \(\kappa\simeq3.15\) for the triplet and \(\kappa\simeq2.46\) for the singlet example.

The value of $\kappa$ and its near independence of the peak follow from the
spectral form of $P(\tau)$ (Appendix~\ref{app:delay}). If finite $U$ shifts the transition
frequencies $\Delta_{kl}=E^{(\pm)}_k-E^{(\pm)}_l$ of Eq.~\eqref{eamp} by $\delta\Delta_{kl}$,
the maximum of $P$ near a peak with aligned phases moves to
\begin{equation}
  \frac{\tau_{\rm Hub}}{\tau_{\rm Heis}}-1
  \simeq-\frac{\sum_{k,l}p_kp_l\,\Delta_{kl}\,\delta\Delta_{kl}}{\sum_{k,l}p_kp_l\,\Delta_{kl}^2},
  \label{eq:kappa}
\end{equation}
a weighted average of the relative frequency shifts that does not depend on the peak.
Equation~\eqref{eq:kappa} gives $\kappa=3.150$ for the triplet and $2.459$ for the singlet,
in agreement with the late peaks; the early peaks, whose phases are not perfectly aligned,
differ by less than $0.004$. Re-timing removes only this average shift. The non-uniform part
of $\delta\Delta_{kl}$ leaves an infidelity that grows as $[\tau\,(t_v/U)^2]^2$ ($0.047$ for
the $535\pi/2$ peak at $U/t_v=150$), so a fixed error requires $U/t_v\propto\sqrt{\tau}$,
which is why the late peaks need much stronger interactions.

The $O(t_b^4/U^3)$ terms of Eq.~\eqref{eq:h_eff} reduce the effective exchange (on a chain,
$J_b=t_b^2/U-4t_b^4/U^3$ in the normalization of
Eq.~\eqref{eq:parameter-map}~\cite{takahashi1977,macdonald1988}), so the magnon moves more slowly.
Figure~\ref{fig:hub-transfer}(f) compares readout at the Heisenberg time with readout at the
delayed time:
\begin{itemize}
\item Measured at the Heisenberg time, the triplet $3\pi$ peak falls to $0.48$ at
  $U/t_v\approx17$, the triplet $535\pi/2$ peak to $0.40$ at $U/t_v=150$, and the singlet
  peaks at $27\pi/2$ and $203\pi$ to $0.77$ at $U/t_v\approx59$ and $0.69$ at $U/t_v=200$.
\item At the corrected time the same peaks give $0.94$, $0.95$, $0.95$ and $0.92$, essentially
  the best values of Fig.~\ref{fig:hub-transfer}(d).
\end{itemize}

Equation~\eqref{eq:delay} is therefore the time map needed to carry the Heisenberg protocol over to
the Hubbard model; because $\kappa$ does not depend on the peak, the same map applies to all high-fidelity peaks of a sector.
For pretty good transfer, finite $U$ has a practical cost: the $535\pi/2$ and $203\pi$ peaks
recover their Heisenberg values only for $U/t_v$ of several hundred, where the exchange
$v=t_v^2/U$ is hundreds of times slower than the hopping. In a Hubbard realization the early peaks, which reach $0.94$
at $U/t_v\approx17$ after re-timing, are the relevant ones.

\section{\label{sec:conclusion}Conclusion and Outlook}

We have studied quantum-state transfer in the 16-site poly-bi-[8]-annulenylene network using Heisenberg and Hubbard Hamiltonians. In the Heisenberg model, the reflection symmetry splits the dynamics into triplet and singlet sectors with closed-form spectra. Neither sector admits perfect transfer. Couplings chosen from Pythagorean triples, however, give pretty good transfer of both states, with arrival times scaling as $\tau_\epsilon\sim\epsilon^{-1}$. Because the rungs only shift the triplet block by a constant, the triplet transfer is independent of $r$, while the singlet transfer is not. The rung coupling can therefore enhance or suppress singlet transfer without affecting the triplet and without time-dependent control. The rungs also make the triplet transfer more robust to coupling disorder than a single SSH-type chain.

In the Hubbard model, the Heisenberg dynamics is recovered in the strong-interaction limit, while finite interactions affect the transfer in two ways. First, the high-fidelity transfer peaks are delayed, $\tau_{\rm Hub}\simeq \tau_{\rm Heis}[1+\kappa\,(t_v/U)^2]$, where $\kappa$ is a spectral average of the finite-$U$ frequency shifts and is therefore the same for every high-fidelity peak of a sector, with $\kappa\approx3.15$ for the triplet and $2.46$ for the singlet example.
 Read out at the Heisenberg time, the transfer drops sharply, so this re-timing is essential. Second, even at the corrected time, the peak fidelity remains below its Heisenberg value until the interaction is strong, since the remaining loss grows as $[\tau_{\rm Heis}(t_v/U)^2]^2$. The early triplet peak reaches its Heisenberg value only from $U/t_v\approx 30$ on, and the long-time pretty-good-transfer peaks only for $U/t_v$ of several hundred, where they are also the most sensitive to disorder. In both models, the early peaks are therefore the practically relevant ones.

These results provide a starting point for symmetry-controlled state transfer in molecular-inspired quantum networks. Such networks could serve as internal state-transfer components in emitter-based quantum nodes, connecting quantum memories, processing elements and spin--photon interfaces. Establishing this will require further study of entanglement generation and preservation, decoherence, coupling to optical emitters, and the fidelity of spin--photon conversion.

\begin{acknowledgments}
This work was supported by the U.S. Department of Energy (DOE), Office of Basic Energy Sciences, under Grant No.~DE-SC0026309; and the US Department of
Energy, the Office of Science through the Quantum Science Center (QSC), a National Quantum Information
Science Research Center.  We also
acknowledge support from the Goodnight Distinguished Chair in Quantum Computing at North Carolina State University.
\end{acknowledgments}

\appendix

\section{Rung couplings and the triplet sector of two-leg ladders}
\label{app:rindep}

We show that, in the single-excitation sector of any two-leg ladder,
the rungs shift the triplet block by a multiple of the identity,
whatever their number, positions and strengths. Number the $2M$ sites
$j=1,\dots,M$ along one leg and $\bar\jmath\equiv 2M+1-j$ along the
other, and let $\Pi$ be the reflection $j\leftrightarrow\bar\jmath$.
We write
\begin{equation}
\label{eHrung}
H=H_0+H_{\rm rung},\qquad
H_{\rm rung}=-\sum_{j=1}^{M} r_j\,\bm{\sigma}_j\cdot\bm{\sigma}_{\bar\jmath},
\end{equation}
with $r_j=0$ where there is no rung, and require only that $H_0$
conserve $\sum_i\sigma_i^z$ and satisfy $\Pi H_0\Pi=H_0$; the leg
couplings are otherwise arbitrary. Poly-bi-[8]-annulenylene has
$M=4N$, $H_0$ given by the $v$ and $w$ bonds of Eq.~\eqref{eq:HN}, and
$r_j=r$ for $j\equiv 2,3\pmod 4$, $r_j=0$ otherwise.

In the single-excitation sector,
$\bm{\sigma}_j\cdot\bm{\sigma}_{\bar\jmath}$ has diagonal element $-1$ on
$|j\rangle$ and $|\bar\jmath\rangle$ and $+1$ on every other state, and
couples $|j\rangle$ to $|\bar\jmath\rangle$ with matrix element $2$.
Hence, for $k=1,\dots,M$,
\begin{align}
\label{erungelem}
\langle k|H_{\rm rung}|k\rangle&=\langle\bar k|H_{\rm rung}|\bar k\rangle
=r_k-\sum_{j\neq k}r_j,\nonumber\\
\langle k|H_{\rm rung}|\bar k\rangle&=-2r_k ,
\end{align}
and all other matrix elements vanish. In the basis
$|k_\pm\rangle=(|k\rangle\pm|\bar k\rangle)/\sqrt2$, which is
Eq.~\eqref{eq:pair-states} for $M=4N$, $H_{\rm rung}$ is therefore
diagonal, with
\begin{equation}
\label{erung}
\langle k_\pm|H_{\rm rung}|k_\pm\rangle
=r_k-\sum_{j\neq k}r_j\mp 2r_k
=\begin{cases}
-R & \text{(triplet)},\\[2pt]
4r_k-R & \text{(singlet)},
\end{cases}
\end{equation}
where $R=\sum_j r_j$. Physically,
$\bm{\sigma}_j\cdot\bm{\sigma}_{\bar\jmath}=+1$ whenever the pair
$(j,\bar\jmath)$ is polarized or in its triplet state, and in the
triplet sector this holds for every pair.

Since $\Pi$ commutes with $H_0$ and with every rung term, the two
sectors decouple (Sec.~\ref{sec:basis}), and the triplet block is
\begin{equation}
\label{ehtr}
h^{(+)}(\{r_j\})=h^{(+)}_0-R\,I_M ,
\end{equation}
with $h^{(+)}_0$ the triplet block of $H_0$. The rungs thus change
triplet amplitudes only by the global phase $e^{iR\tau}$, so every triplet
transfer probability [Eq.~\eqref{eamp}], between the ends or between
bulk pairs, is independent of the rung couplings. In the singlet sector the rungs instead add the site-dependent
potential $4r_k$, which changes both the eigenvalues and the
eigenvectors [Eq.~\eqref{eenersing}].

\section{Explicit transfer probability}
\label{app:ptexplicit}
 
For $(v,w)=(4,3)$, inserting the eigenvalues of Eq.~\eqref{eavtrip431}
and the corresponding eigenvectors of Eq.~\eqref{emh8t}
into Eq.~\eqref{eamp} gives $P^{(+)}(\tau)$. It is the same for
every $r$, since $r$ only shifts all triplet eigenvalues by a constant:
\begin{widetext}
\begin{equation}
\label{eptv4w3}
\begin{aligned}
P^{(+)}(\tau)
= \frac{1}{269600}\Bigg\{&
337\Big[
40\cos(4\tau)+10\cos(8\tau)+40\cos(12\tau)
+25\cos(16\tau)+16\cos(20\tau)+10\cos(24\tau)
\Big]
\\
&+200\alpha_+^2\cos(4\alpha_- \tau)
 +200\alpha_-^2\cos(4\alpha_+ \tau)
\\
&+50(337+\sqrt{337})
  \cos\!\left(2\sqrt{50-2\sqrt{337}}\,\tau\right)
 +50(337-\sqrt{337})
  \cos\!\left(2\sqrt{50+2\sqrt{337}}\,\tau\right)
\\
&+\sum_{\sigma=\pm}
\Bigg[
-16850\cos(2\alpha_\sigma \tau)
\Big(
\cos(2\tau)+2\cos(10\tau)+\cos(14\tau)
\Big)
\\
&\hspace{3.2cm}
+10\beta_\sigma\sin(2\alpha_\sigma \tau)
\Big(
5\sin(2\tau)-6\sin(10\tau)-5\sin(14\tau)
\Big)
\Bigg]
+43583
\Bigg\},
\end{aligned}
\end{equation}
\end{widetext}
where $\alpha_\pm=\sqrt{25\pm12\sqrt2}$ and
$\beta_\pm=\sqrt{337(137\pm96\sqrt{2})}$. The frequencies are the
eigenvalue differences of Eq.~\eqref{eavtrip431}: the integers
$4,8,\dots,24$, the combinations $2\alpha_\pm\pm2$, $2\alpha_\pm\pm10$,
$2\alpha_\pm\pm14$, $4\alpha_\pm$, and
$2(\alpha_+\mp\alpha_-)=2\sqrt{50\mp2\sqrt{337}}$. All of them are integer
combinations of $1$, $\alpha_+$ and $\alpha_-$, so $K_I=2$.

\section{\label{app:otoc}OTOC and transfer}

OTOCs are commonly defined as four-point correlation functions
\begin{equation}
F_{WV}(\tau)=\left\langle W^\dagger(\tau)V^\dagger W(\tau)V\right\rangle,
\quad W(\tau)=e^{iH\tau}We^{-iH\tau},
\end{equation}
and quantify the spreading of initially local operators under
many-body dynamics \cite{xuScramblingDynamicsOutofTimeOrdered2024a}.
Fidelity OTOCs \cite{lewis-swanUnifyingScramblingThermalization2019a}
take $V$ to be the projector on the initial state. We make the same
choice, $V=|a\rangle\langle a|$, take $W=|b\rangle\langle b|$ to be the
projector on the target state, and evaluate the OTOC in $|a\rangle$.
Since $V|a\rangle=|a\rangle$ and $V$ has rank one,
\begin{align}
\label{eq:otoc}
F(\tau)&=\langle a|W(\tau)VW(\tau)V|a\rangle\nonumber\\&=\langle a|W(\tau)|a\rangle^2
\nonumber\\&=\left|\langle b|e^{-iH\tau}|a\rangle\right|^4\nonumber\\&=P_{a\to b}(\tau)^2 .
\end{align}
For $|a\rangle=|1\rangle$ and $|b\rangle=|9\rangle$ this is the density
OTOC $\langle 1|n_9(\tau)n_1n_9(\tau)n_1|1\rangle$, since in the
single-excitation sector $n_i=|i\rangle\langle i|$. The squared commutator likewise reduces to
\[\langle a|[W(\tau),V]^\dagger[W(\tau),V]|a\rangle
=P_{a\to b}(\tau)\,[1-P_{a\to b}(\tau)].\] Both identities hold for any
Hamiltonian, including the tight-binding and Hubbard models. For this
choice of operators the OTOC is therefore not an independent transport
observable: it encodes the same amplitude as the transfer probability.
This has a practical side: OTOCs are usually measured with time-reversal
or interferometric protocols \cite{swingleMeasuringScramblingQuantum2016,garttnerMeasuringOutoftimeorderCorrelations2017}, whereas
$P_{a\to b}(\tau)$ is a direct occupation measurement at the target, so for
state transfer the OTOC offers no experimental advantage. It also means
that the time dependence of such projector OTOCs reflects only the
propagation of the excitation, not scrambling.

\section{\label{app:delay}Delay of the transfer peaks at finite \texorpdfstring{$U$}{U}}

This appendix derives Eq.~\eqref{eq:kappa} and the infidelity that remains after
re-timing. As in the definition of $\kappa$ [Fig.~\ref{fig:hub-transfer}(e)], we keep the eight
bound states in Eq.~\eqref{eq:hub-amp}. Let $\varepsilon_k$ be their energies, labeled so that
$-\varepsilon_k-J_{\rm tot}\to E_k$ for $U\to\infty$, where $E_k\equiv E^{(\pm)}_k$ are the
Heisenberg energies of Eq.~\eqref{eenertri} or~\eqref{eenersing}. We write
\begin{equation}
  -\varepsilon_k-J_{\rm tot}=E_k+\delta E_k,\qquad
  \delta\Delta_{kl}=\delta E_k-\delta E_l,
\end{equation}
with $\delta E_k\propto(t_v/U)^2$ at large $U$. The reflection and shift do not change the
transfer probability (Sec.~\ref{sec:hubbard-model}), and we take the parities $s_k$ and the
weights $p_k$ from the Heisenberg model (see the end of this appendix). With
$\Delta_{kl}=E_k-E_l$ and $\sum_kp_k=1$,
\begin{equation}
  1-P(\tau)=\sum_{k,l}p_kp_l\big\{1-s_ks_l\cos[(\Delta_{kl}+\delta\Delta_{kl})\tau]\big\}.
  \label{eq:app-infid}
\end{equation}

We first assume that the phases are perfectly aligned at the Heisenberg peak $\tau_{\rm Heis}$,
i.e.\ that Eq.~\eqref{ephase} holds there. No peak satisfies this exactly (Sec.~\ref{sec:pt});
the corrections are given at the end of this appendix. Then
$s_ks_l\cos(\Delta_{kl}\tau_{\rm Heis})=1$ and $\sin(\Delta_{kl}\tau_{\rm Heis})=0$ for all
$k,l$. Near this peak, at $\tau=\tau_{\rm Heis}+\delta\tau$, the argument of each cosine is
$\Delta_{kl}\tau_{\rm Heis}+x_{kl}$, with
\begin{equation}
  x_{kl}=\Delta_{kl}\,\delta\tau+\delta\Delta_{kl}\,\tau_{\rm Heis}
\end{equation}
up to the higher-order term $\delta\Delta_{kl}\,\delta\tau$. Each term of
Eq.~\eqref{eq:app-infid} then becomes $1-\cos x_{kl}$, and to second order
\begin{equation}
  \begin{aligned}
    1-P&\simeq\frac12\sum_{k,l}p_kp_l\,x_{kl}^2\\
    &=\frac12\big(S_2\,\delta\tau^2+2S_1\,\tau_{\rm Heis}\,\delta\tau+S_0\,\tau_{\rm Heis}^2\big),
  \end{aligned}
  \label{eq:app-quadratic}
\end{equation}
where
\begin{equation}
  \begin{gathered}
    S_2=\sum_{k,l}p_kp_l\,\Delta_{kl}^2,\qquad
    S_0=\sum_{k,l}p_kp_l\,\delta\Delta_{kl}^2,\\
    S_1=\sum_{k,l}p_kp_l\,\Delta_{kl}\,\delta\Delta_{kl}.
  \end{gathered}
\end{equation}
There is no term linear in $x_{kl}$, because $\sin(\Delta_{kl}\tau_{\rm Heis})=0$. The expansion
requires only that the $x_{kl}$ be small near the new peak. The phase errors
$\delta\Delta_{kl}\tau_{\rm Heis}$ at the old peak need not be: for the late peaks they are of
order one where the transfer at $\tau_{\rm Heis}$ has collapsed
[Fig.~\ref{fig:hub-transfer}(f)].

Minimizing Eq.~\eqref{eq:app-quadratic} over $\delta\tau$ gives the peak time
$\tau_{\rm Hub}=\tau_{\rm Heis}+\delta\tau$ with $\delta\tau/\tau_{\rm Heis}=-S_1/S_2$, which is
Eq.~\eqref{eq:kappa}. The peak time $\tau_{\rm Heis}$ has dropped out, so all aligned peaks of
a sector are delayed by the same fraction. The ratio $-S_1/S_2$ is the average of
$-\delta\Delta_{kl}/\Delta_{kl}$ with weights $p_kp_l\Delta_{kl}^2$; for a uniform rescaling,
$\delta\Delta_{kl}=-\lambda\Delta_{kl}$, it gives $\delta\tau/\tau_{\rm Heis}=\lambda$ to leading order.
Since $\delta\Delta_{kl}\propto(t_v/U)^2$, Eq.~\eqref{eq:delay} follows with
$\kappa=-\lim_{U\to\infty}(U/t_v)^2S_1/S_2$.

Inserting the optimal $\delta\tau$ into Eq.~\eqref{eq:app-quadratic} gives the infidelity at
$\tau_{\rm Hub}$,
\begin{equation}
  1-P(\tau_{\rm Hub})\simeq\frac{\tau_{\rm Heis}^2}{2}\left(S_0-\frac{S_1^2}{S_2}\right).
  \label{eq:app-residual}
\end{equation}
By the Cauchy--Schwarz inequality, $S_1^2\le S_0S_2$, with equality only for a uniform
rescaling of the frequencies. Equation~\eqref{eq:app-residual} is therefore the part of the
frequency shifts that re-timing cannot remove. Since $S_0-S_1^2/S_2\propto(t_v/U)^4$, it grows
as $[\tau_{\rm Heis}(t_v/U)^2]^2$. For the $535\pi/2$ peak at $U/t_v=150$ it gives $0.0473$; the
exact loss relative to the Heisenberg peak, at the corrected time of
Fig.~\ref{fig:hub-transfer}(f), is also $0.0473$. For the $203\pi$ peak at $U/t_v=200$ the two
values are $0.087$ and $0.082$.

Real peaks are not perfectly aligned, and the derivation neglects three effects. (i)~At the
Heisenberg maximum the term linear in $\delta\tau$ still vanishes, since its coefficient is
proportional to $dP/d\tau$. However, the quadratic terms acquire the factor
$s_ks_l\cos(\Delta_{kl}\tau_{\rm Heis})$, and the change of the weights $p_k$ at finite $U$, of
relative order $(t_v/U)^2$, now shifts the peak at that order. Together these change $\kappa$ by
less than $0.004$ for the early peaks and $0.001$ for the late ones. (ii)~Misalignment also adds
a loss linear in $\delta\Delta_{kl}$. It decreases only as $\tau_{\rm Heis}(t_v/U)^2$ and
accounts for most of the difference between Eq.~\eqref{eq:app-residual} and the exact loss below
about $1\%$. (iii)~The initial state has a weight of order $(t_v/U)^2$ on doublon--holon states,
which lowers $P$ by a similar amount, negligible for the late peaks; its fast oscillations are
excluded from $\kappa$, which is defined by the peak of the bound-band amplitude.

\bibliography{apssamp}% Produces the bibliography via BibTeX.

\end{document}